\documentclass{ijuc}

\usepackage[pdftex]{graphics}
\usepackage{amssymb}
\usepackage{subfig}
\usepackage{xurl}
\usepackage{xcolor}

\usepackage{hyperref}
\hypersetup{
    colorlinks=true,
    linkcolor=blue,
    filecolor=black,      
    urlcolor=cyan,
    citecolor=black,
    pdfpagemode=FullScreen,
    }

\usepackage{amsmath, graphics, setspace}
\usepackage{booktabs} 
\usepackage[flushleft]{threeparttable} 
\usepackage{credits}

\newcommand{\mathsym}[1]{}
\newcommand{\unicode}[1]{}

\begin{document}

\title{Information Encoding and Decoding in In-Vitro Neural Networks on Micro Electrode Arrays through Stimulation Timing}

\author{Trym A. E. Lindell\inst{1}\email{trym.a.e.lindell@gmail.com}
\and Ola H. Ramstad \inst{1} 
\and Ioanna Sandvig\inst{2}
\and Axel Sandvig \inst{2}
\and Stefano Nichele \inst{3}
}

\institute{Department of Computer Science, Oslo Metropolitan University, Norway
\and
Department of Neuromedicine and Movement Science, Norwegian University of Science and Technology, Norway
\and
Department of Computer Science and Communication, Østfold University College, Norway
}

\def\received{}

\maketitle

\begin{abstract}
A primary challenge in utilizing in-vitro biological neural networks for computations is finding good encoding and decoding schemes for inputting and decoding data to and from the networks. Furthermore, identifying the optimal parameter settings for a given combination of encoding and decoding schemes adds additional complexity to this challenge. 
In this study we explore stimulation timing as an encoding method, i.e. we encode information as the delay between stimulation pulses and identify the bounds and acuity of stimulation timings which produce linearly separable spike responses. We also examine the optimal readout parameters for a linear decoder in the form of epoch length, time bin size and epoch offset. 
Our results suggest that stimulation timings between 36 and 436ms may be optimal for encoding and that different combinations of readout parameters may be optimal at different parts of the evoked spike response. 

\end{abstract}


\section{Introduction}

\label{sec:Intro}
Cutting-edge AI systems require vast amounts of computing power and, consequently, significant energy consumption. A prime example is the 2023 large language model GPT-4, which, during training may have cost more than 100 million USD according to a statement by OpenAI's CEO Sam Altman \cite{knight_2023}. While the performance of such models is impressive, comparable results are observed in biological networks at a fraction of the energy cost. As an example, the human brain only uses approximately 20 watts of energy\cite{balasubramanian2021brain}. Moreover, advances in neuroengineering now enable relatively easy access to biological neural networks for  bio-computing \cite{kagan2023technology}. 
By harnessing biological computing to address problems currently delegated to Artificial Intelligence (AI) systems, we could thus significantly reduce the burden on our planet and, in turn, benefit future society.

Several studies have already explored this potential solution, with researchers developing biological computational systems capable of solving simple tasks, such as playing Pong \cite{kagan2022vitro}, adaptive flight control \cite{demarse2005adaptive} and controlling basic robots/animats \cite{demarse2001neurally, martinoia2004towards, karniel2005computational, bakkum2007embodying, novellino2007connecting, xydas2008architecture, warwick2010controlling,warwick2011experiments, tessadori2012modular,li2015application, aaser2017towards, takahashi2016reservoir, masumori2018autonomous} (for overview of robotics applications see: \cite{chen2023overview}. To achieve such results, one approach involves treating the biological substrate as a reservoir within a reservoir computing paradigm \cite{aaser2017towards}. This involves inputting data into the biological substrate, allowing it to perform computations on said data, and then extracting these computations using a linear machine learning decoder to produce the desired outputs. Although such systems still utilize a traditional hardware component, the required computing power is significantly lower than that of large, deep models, as only the simple linear decoder is trained. Another benefit specific to utilizing in-vitro neural networks in this framework is that reservoir computing does not require training of the reservoir \cite{maass2002real, jaeger2001echo} (although closed loop systems allowing for training can be used \cite{aaser2017towards}). Thus, no feedback signals to the networks are required, making experimentation and use far simpler.

Several challenges arise when creating such systems; What is the optimal method of encoding data into the biological network? What are the bounds of such encoding scheme and what is its acuity? What is the best method of re-translating the recorded spikes to input for a linear decoder, i.e. how is the relevant information encoded in the network activity?  Furthermore, are the networks stable enough to be useful for practical applications without needing constant re-training?

In this paper we tackle these challenges for biological computing systems using in-vitro neural networks as their computing substrate. We target the timing between electrical stimulations as our encoding method, i.e we investigate the range of delays between two stimulation pulses that can be used to encode input values, and explore three key features of this method. Firstly we aim to identify the upper bound of separable stimulation timing (what is the maximal delays that can be used), secondly we explore the lower bound (what is the minimal delay that can be used) and thirdly we explore the acuity of this separability (how close can two delays be and still be linearly separated). Furthermore, we explore how the network encodes its computations over time, how this affects optimal decoder-readout choices and the stability of the networks on a day to day timeline. The goal is thus to to assess the potential and optimal parameters for encoding information through stimulation timing for later use in reservoir computing, and the optimal parameters for decoding information input through this method from the spiking activity produced by the in-vitro neural reservoirs. 
\newline \newline

The paper is organized as follows. We first provide relevant background and justification for the computing system and encoding/decoding method. 
This is followed by general methods concerning the computing system and the experimental methods. In the analysis and results section we detail the motivation, methods and results aimed at answering each separate aspect of the challenges described above and briefly discuss the results. Finally we summarize and discuss the overall results in the summary discussion section. 

\section{Background}
\label{sec:Backg}
The proposed computing system explored in this paper consists of a biological neural network and a Micro Electrode Array (MEA) on which the neurons are grown. The paradigm for which the encoding and decoding methods are being assessed is reservoir computing. In this section we first introduce reservoir computing, followed by a description of the in-vitro neural networks and MEA. Lastly we provide a short review on related and relevant literature to reservoir computing using in-vitro neural networks and identify the knowledge gaps this study aims to fill.

\subsection{Reservoir Computing}
\label{sec:resC}
Reservoir computing is a field within artificial intelligence where the natural computations of a system can be harnessed without a large training overhead. The general framework arose independently multiples times, initially through the works of \cite{kirby1990neurodynamics, kirby1991context} and later in the work of \cite{schomaker1991simulation, schomaker1992neural, dominey1995complex, jaeger2001echo, maass2002real}. The most well known examples comes from \cite{jaeger2001echo} in the form of Echo State Networks (ESN) and \cite{maass2002real} in the form of the Liquid State Machine (LSM). ESN are value continuous, time-discrete artificial neural networks which rely on the echo state, meaning that the states of the network contracts over time, or in other words, it gradually forgets its previous inputs. The LSM was originally formulated as a general computing framework for mapping time series with fading memory to time series with fading memory. This was expanded to apply to spiking neural networks (value discreet, time continuous systems) with the first implementation of a LSM using artificial spiking neural networks. Although a general framework, LSM now often refers specifically to reservoir computing systems using spiking neurons. LSMs thus translates directly to in-vitro neural networks as these also consists of spiking neurons. Both system types shares a range of properties and is generally unified under the term reservoir computing \cite{verstraeten2005reservoir, verstraeten2007experimental}

\begin{figure}[!htb]
    \centering{}
    \includegraphics[width = 0.7\linewidth]{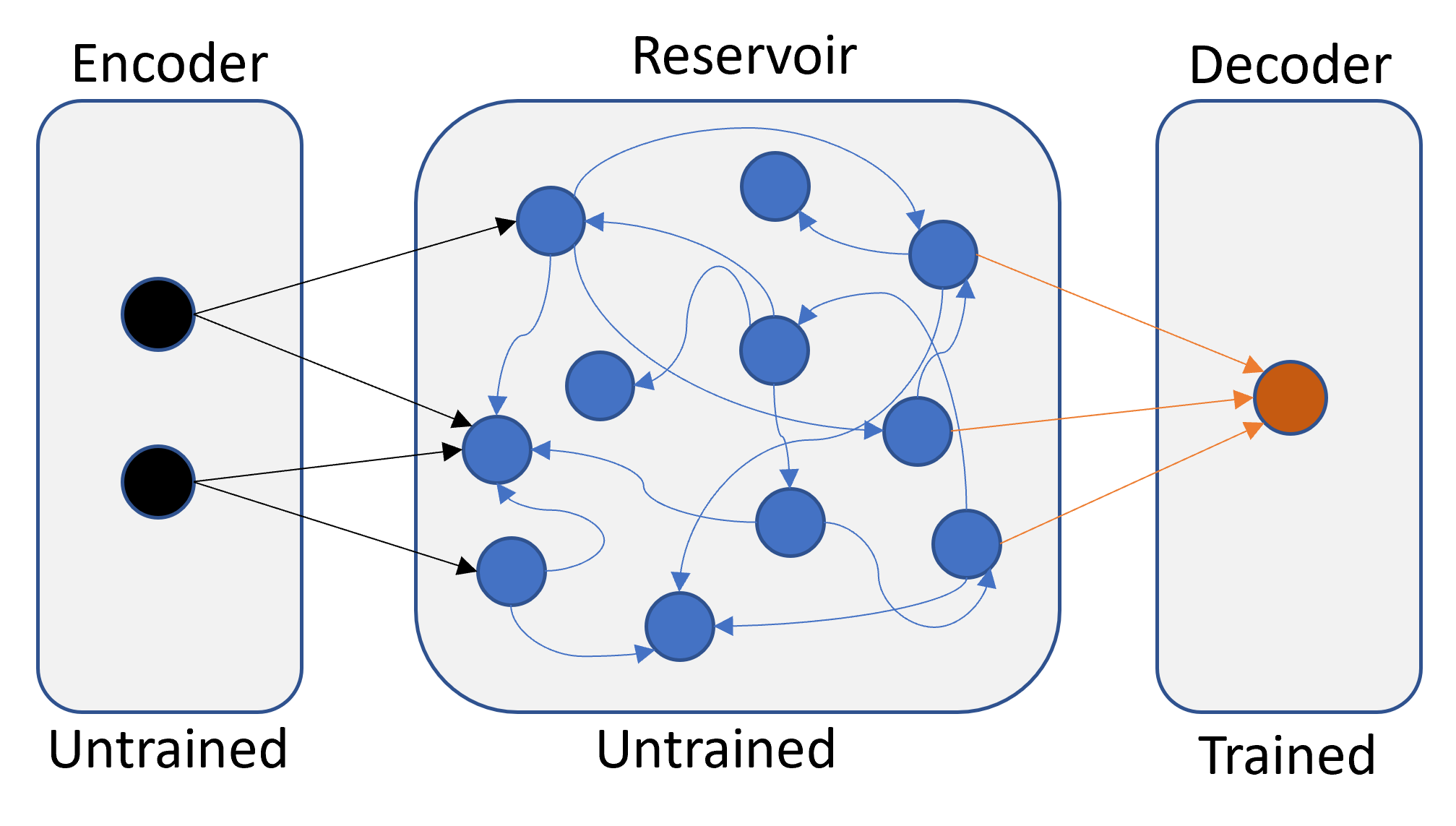}
    \caption{Components of minimal reservoir computing system. Input is given through the encoder into the reservoir, in this case a recurrent neural network. The input propagates through the reservoir network and the decoder reads out the reservoir states and translates them to the desired output after training.}
    \label{fig:ResComp}
\end{figure}

A typical reservoir computing system can be divided into three main components, an encoder, a reservoir and a decoder (often called readout), visualized in Figure \ref{fig:ResComp}. The defining component is the reservoir. This consists of a dynamical system, be it an artificial neural network \cite{jaeger2001echo, maass2002real, verstraeten2007experimental}, a bucket of water \cite{fernando2003pattern}, a cat \cite{nikolic2006temporal}, or in our case in-vitro neural networks. 

The reservoir is generally not trained but receives input from the encoder which are propagated through the reservoir. The encoder translates the input data into a format the reservoir can compute on. In the case of the bucket of water study \cite{fernando2003pattern} this consisted of encoding digital binary values into the movement of motors embedded in the water. The decoder reads out the states of the reservoir and translates them into the desired output using a machine learning system. Notably the decoder can be subdivided into a readout component that records the states of the reservoir, transforming it to input to the machine learning component and the machine learning component which translates this input to the desired output. Again returning to the bucket of water in \cite{fernando2003pattern}, the readout consisted of a video camera reading out the wave patterns caused by the movement of the encoder motors and the machine learning component was a 50- neuron parallel perceptron system using the p-delta rule. Given this setup the parallel perceptrons could learn to classify the input based on the wave pattern recorded through the video camera. 
The machine learning component is often some linear method as this both reduces training costs, as only the readout is trained, and allows us to test if the reservoir performed relevant computations by inputting a non-linear problem and assessing if the linear machine learning component can solve it based on the reservoir states. If this is the case we say that the reservoir has linearlized the input problem.

\subsection{In-Vitro Neural Networks on Micro Electrode Arrays}
\label{sec:inVitNN}
\begin{figure}[!h]
    \centering{}
    \includegraphics[width = 1\linewidth]{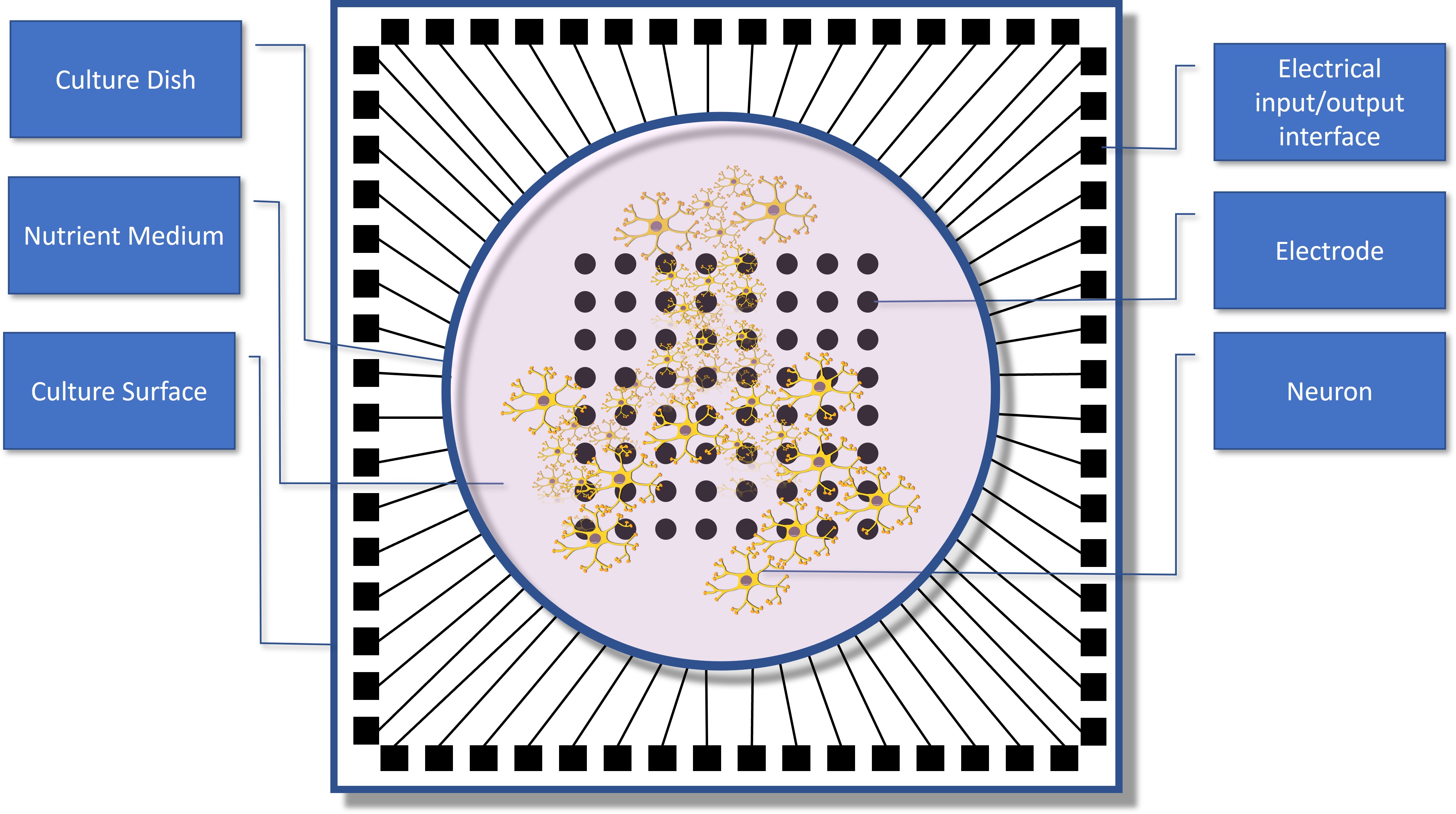}
    \caption{Schematic figure of in-vitro neural network on Micro Electrode Array (MEA).}
    \label{fig:SchematicMEA}
\end{figure}

In our experiments we have used biological neural networks derived from rat cortical neurons, grown in-vitro as our reservoir substrate. These consists mainly of two cell types: cortical neurons and astrocytes. A schematic figure of a MEA device is shown in Figure \ref{fig:SchematicMEA}.

\begin{figure}[!htb]
    \centering{}
    \includegraphics[width = 0.7\linewidth]{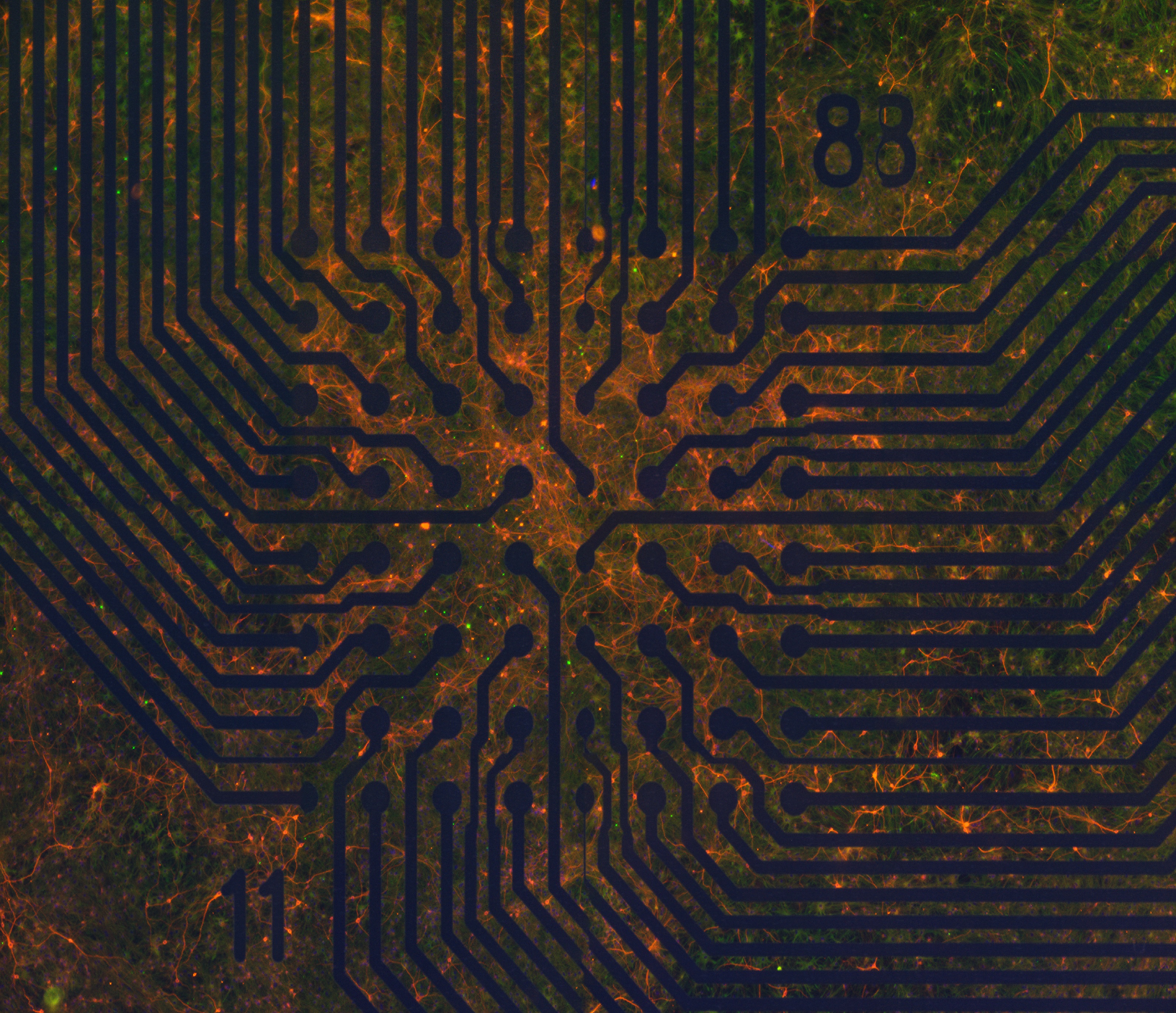}
    \caption{Fluorescent microscopy of an in-vitro neural network on an MEA interface after immunostaining : neurons (Neurofilament heavy; red), astrocytes  (GFAP;green); cell nuclei (Hoechst; blue).}
    \label{fig:InVitroMEA}
\end{figure}

In Figure \ref{fig:InVitroMEA} we show an in-vitro neural network grown on a Micro Electrode Array. The neurons are stained red, while the astrocytes are colored green. Cell cores are colored blue and electrodes are seen as black circles. Note, that these images are not from the networks in this study, which are denser. 

When the neurons are added to their container they are initially disconnected but rapidly self organize into a 2D network. As they develop they begin producing spontaneous spikes and at later stages may elicit network wide bursts\cite{rutten2001activity}. Spiking activity can be measured as electrical waves over electrodes embedded in the bottom of the container. Containers with this interface are named Micro Electrode Arrays (MEA) and allows for both measurement of electrical activity and for electrical stimulation of the cells. Such stimulation can activate the neurons causing them to produce spikes which can spread through the network and may thus serve as an input channel for information. 

\begin{figure}[ht]
    \centering{}
    \includegraphics[width = 0.9\linewidth]{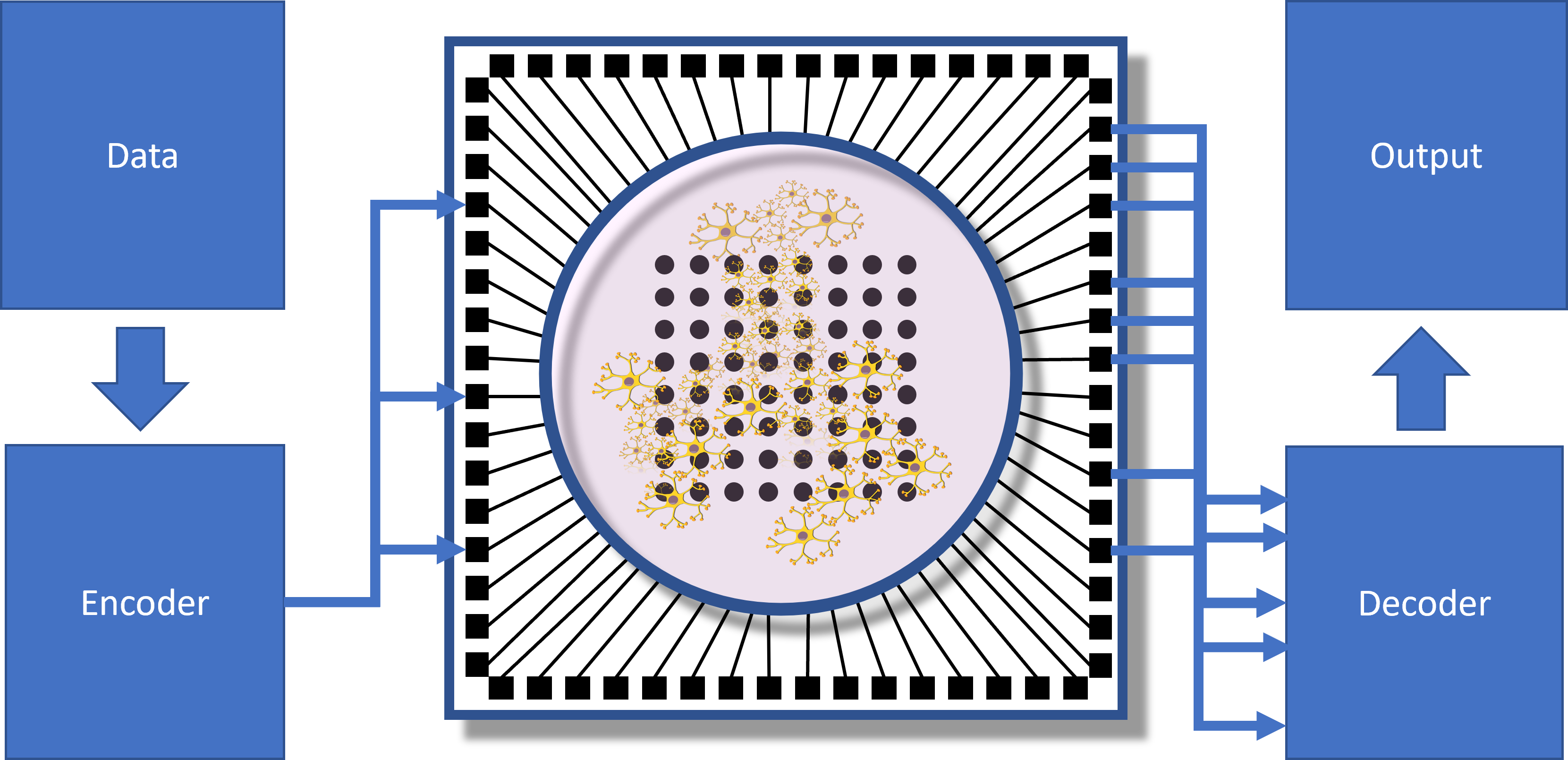}
    \caption{Schematic figure of in-vitro neural network on Micro Electrode Array (MEA) as a reservoir computing system}
    \label{fig:SchematicMEAReservoir}
\end{figure}

Since the MEA allows for bi-directional communication with the in-vitro neurons it serves as both a component in the encoder and as the readout of the decoder. The neural network itself serves as a reservoir which computes on the input given as electrical stimulations and produces spike patterns recorded through the MEA which is then given to the machine learning component of the decoder, thus forming a reservoir computing system. We show a schematic Figure of the in-vitro neural network - MEA - reservoir computing system in Figure \ref{fig:SchematicMEAReservoir}.

\subsection{Reservoir Computing With in-Vitro Neural Networks}
\label{sec:RCinVit}
According to the LSM framework a reservoir (liquid) must have the separation property to be able to function as a reservoir \cite{maass2002real}. This means that for significantly different inputs, the reservoir produces separable states. 
Since different substrates have fundamentally different dynamics, one thus needs to establish an encoding method that translates ones data into a format that can evoke separable dynamics within the substrate. For a bucket of water, mechanical perturbations works well as it produces waves on the water's surface. For in-vitro neural networks, such input is unlikely to produce the desired response. 

Since neural spike propagation is an electrical phenomenon, electrical encoding methods is the natural option for inputting information into a biological neural circuit. Electrical encoding can however be done in many different ways. One may for example vary voltage, current, frequency or the timing between stimulation pulses. Furthermore, one may multiplex many of these options. The possible input space is high dimensional. 

Multiple studies have explored different input encodings that produce separable network states (although not always with this explicit goal): \cite{wagenaar2004effective, wagenaar2005controlling, baljon2009interaction, shew2009neuronal, hafizovic2007cmos, dockendorf2009liquid, george2014input, pimashkin2016selectivity, george2018random, ortman2011input, dranias2013short}, which demonstrates that in-vitro neural networks in theory can be used as reservoirs. However, a detailed exploration of the separability of each input method is lacking. 
Specifically, assessment of an encoding method should investigate relevant features that will limit how input data can be translated to encoded input. 

If, for example, frequency is used to encode input data we would need to ascertain the limits of the lower and upper bounds of frequency the network can separate as well as how small variations in frequency can be distinguished. These bounds are essential to create encoding methods as they tell us within what ranges we must bound our input and with what acuity we can encode it. We may for example have an input data range of 0 - 256 but frequency bounds of 10 - 60 hz outside of which the network gives inseparable responses, meaning that the 0 - 256 range must be mapped to the 10 - 60 hz range. Furthermore, if the network only gives separable responses to frequency differences of more than 1 hz and our input range includes all integers between 0 and 256, as would be the case for image pixels, we know that we may need to use a different encoding method or multiplex it with for example a spatial encoding to be capable of computing on the full input range. 

The considerations above illustrate the importance of thoroughly examining various encoding methods, using specific characteristics such as frequency bounds as an example of the constraints that might be encountered. Understanding these constraints is essential in determining how best to translate input data into an encoded form. Such a comprehensive approach may allow us to identify the most effective encoding strategies and may reveal the potential of different methods. 

Out of the possible options mentioned previously, stimulation timing provides a range of benefits over the other methods making it a prime target for investigation. It allows high density information throughput as one can encode information in the timing between just two input pulses, as opposed to over many, as is needed for frequency. It can be multiplexed with many other encoding methods and can thus serve as a good basis for even more information dense signals. Lastly the presence of spike-time dependent plasticity in the brain suggests that timing between two pulses is a natural information code used in the brain which may make it more likely that the networks may compute well on such signals. 

Applied over two electrodes this encoding method has been explored by \cite{shkolnik2003neurally} who showed that timing between two stimulation pulses, delivered to two separate electrodes, can be used to solve boolean logic and control animat robots. They explored timing differences from 0 to 1 second.
Applied to a single MEA electrode, this input methods has been partly explored by \cite[]{baljon2009interaction} who examined how the evoked response from two stimulation pulses interact. They tested both single electrode and two electrode stimulation. For the single electrode protocol they found that the second stimulation did not have much impact on the evoked response for stimulation delays up to 250 ms, suggesting a lower bound for this input method on single electrodes. However, their analysis method would not be able to extract spatio-temporal, high acuity, spike time interactions as they tested single or pooled electrodes with 5ms time bins, which makes this lower bound uncertain. They also tested inter-stimulation intervals up to 1000 ms, finding interactions between the evoked responses which suggests separability at least up to this delay, but leaving the upper bound uncertain.  
An assessment of the separability between the evoked responses between two different delays has also not been established and would be essential as this would limit the network's ability to compute on small differences in stimulation timing.

Another key issue in creating a reservoir computing system using physical reservoirs is how the readout component of the decoder performs its readout, prepossessing and passing to the machine learning component. 
To the best of the authors' knowledge, no studies have explored the optimal settings for extracting the spatio-temporal spike patterns recorded through MEA for the purpose of reservoir computing. However, \cite{wagenaar2004effective} show that the evoked responses from single pulses display different phases consisting of more or less consistent spike times which suggests that information may be encoded in different forms and that this may change over the temporal course of the response. The different phases of this response was used to control the animat in \cite{shkolnik2003neurally} such that the direction of the robot was decoded from the first phase and the magnitude of the direction vector was decoded from the second phase. This suggests that different parameter settings may be optimal for different phases of the response. This issue will require investigation as sub-optimal readout choices may mask the benefits of a particular encoding method. E.g: if stimulation timing produces separable responses at the spike timing level, but we only record rate information we may falsely conclude that stimulation timing is not a good encoding method.

In summary, initial exploration of in-vitro neural networks on MEA have examined a range of encoding methods showing promising results. A natural next step is to conduct targeted investigations that can assesses the relevant features, namely the upper and lower bound of separability and its acuity, with the purpose to create an optimal encoder using stimulation timing. Furthermore, the optimal readout settings for this purpose are unknown, but are likely to involve multiple different information codes and thus requires additional investigation.

\section{Methods}
\label{sec:Meth}
The methods section details the general experimental setup for our experiments. It is divided into two main sections; System methods and Experimental methods. The system methods section details the procedure for establishing the neural networks, the recording system, the spike detection settings and the machine learning protocols. In the experimental methods section we describe the stimulation procedure for the two stimulation experiments we conducted.
Further details concerning the analysis methods is given in the Analysis and Results section.
The code and data used in this paper is available at the digital repository: \url{https://osf.io/yzf47/}.

\subsection{System Methods}
\label{sec:SysMeth}
\subsubsection{Neural Network Cultures}
\label{sec:MethNNCult}
To establish the in-vitro neural networks, rat cortical neurons (A1084002, Thermo-Fisher Scientific) were seeded on Cytoview 6-well plates (Axion Biosystems) at a density of 1,800 cells/mm2 for a total of approx. 190,000 cells per well (cell handling was conducted according to MAN0001574, Thermo-Fisher Scientific). In addition to the rat cortical neurons, a co-culture with rat primary cortical astrocytes was established using concurrent seeding of 180 cells/mm$^2$. Prior to seeding the wells were coated with polyethyleneimine at 0.1\% for 1 hour, (Polysciences), washed with sterile deionized water, left to air dry over-night before a coating of natural mouse laminin (Thermo Fisher Scientific). 

Following seeding the cells were maintained using 95\% Neurobasal Plus, 2\% B27 Plus Supplement (50X), 1\% GlutaMax Supplement, and 2\% PenStrep (all from Thermo Fisher Scientific). The media was exchanged first 24 hours post seeding, then every three days following, with the exception of one feeding at 6 days in vitro (DIV). No experiments were conducted within 24 hours following media changes. 

\subsubsection{MEA Recording System}
\label{sec:MEASys}
The Maestro Pro system recorded at 12.5 khz, using a 3khz Kaiser window for low pass filtering and a 200 hz IIR for high pass filtering. Referencing method was set to "Median" and analog mode setting to "Neural Spikes". The software version for AxIS  and Maestro Pro firmware was 2.0.4.21, Maestro Prop BioCore Version was "4B". CO$_2$ concentration was set to 5.0\% while temperature was set to 37.0 C.

 The micro electrode array type was CytoView MEA plate (M384-tMEA-6B, Axion Biosystems).  Each Cytoview plate consists of 6 wells of 64 PEDOT electrodes with 100 um diameter and 300 um pitch arranged in a square 8x8 array. 
 Networks were maintained in the Maestro Pro incubator continuously following 13 DIV to keep recording conditions consistent, the MEA were only removed for 30 minutes during each media change. 

\subsubsection{Spike and Network Burst Detection}
\label{sec:SpikeDet}
We used the "adaptive threshold crossing" spike detection method with a threshold set to 7 provided by the Axion software. I.e the standard deviation of the continuous electrical recording is computed on an electrode basis and spikes are detected when the signal amplitude reaches a value of 7 times the standard deviation.

The electrode burst detection algorithm was set to inter-spike interval threshold. The maximum inter-spike interval was 100 ms, and the minimum number of spikes was 5. Furthermore, for network burst detection, the minimum number of participating electrodes was set to 25$\%$ (16 electrodes). I.e electrode bursts were detected as a minimum of 5 spikes occurring with less than 100 ms delay between, and networks bursts additionally required spikes to be detected at at least 16 electrodes.

\subsection{Machine learning algorithm}
\label{sec:MLAlg}
In the original liquid state machine the decoder would utilize the high dimensional projection of the input to allow computing equivalences between (potentially) unstable states to produce stable output \cite{maass2002real}. Similarly, standard reservoir computing is often used for time series prediction where linear regression is used to predict future values of an input time series based on the state trajectory of the reservoir\cite{soltani2023echo}. The goal for this paper is however not to create a reservoir computing system per se, but to identify the bounds and acuity of linear separability of different stimulation delays to assess if this encoding method can be used to encode input to a biological neural reservoir. This requires that we identify when the evoked response from different stimulation delays can be linearly separated and when it cannot. Furthermore, we need to assess what readout parameters are optimal for extracting the information contained in the evoked responses. 
Therefore we train separate logistic regression models for combinations of epoch size, time bin size and epoch offset (detailed in section \ref{sec:DecReadInfoCodeMeth}) for the different stimulation delays (detailed in section \ref{sec:AnRes}). The success or failure of the regression models at categorizing a given epoch as belonging to a given stimulation delay thus becomes the measure of linear separability.

For this purpose we used SKlearn's Logistic regression algorithm (version 1.1.3). 
The datasets were split into training and test sets using train\_test\_split from SKlearn with stratify set to True and a 70 - 30 percent train to test split ratio. Training was done using LogisticRegression with settings: penalty = l2, solver = liblinear. 
Since spontaneous bursting activity potentially could affect the evoked responses caused by stimulation if they coincided closely with the cue or probe stimulation the train - test split and training was redone 100 times to average out potential interference. The mean performance is presented in the results. For reproducibility the random state of the train-test split and LogisticRegression was set to the integer value of the iteration (0-99).

\subsection{Experimental Methods}
\label{sec:ExpMeth}
\begin{figure}[!h]
    \centering{}
    \includegraphics[width = 0.6\linewidth]{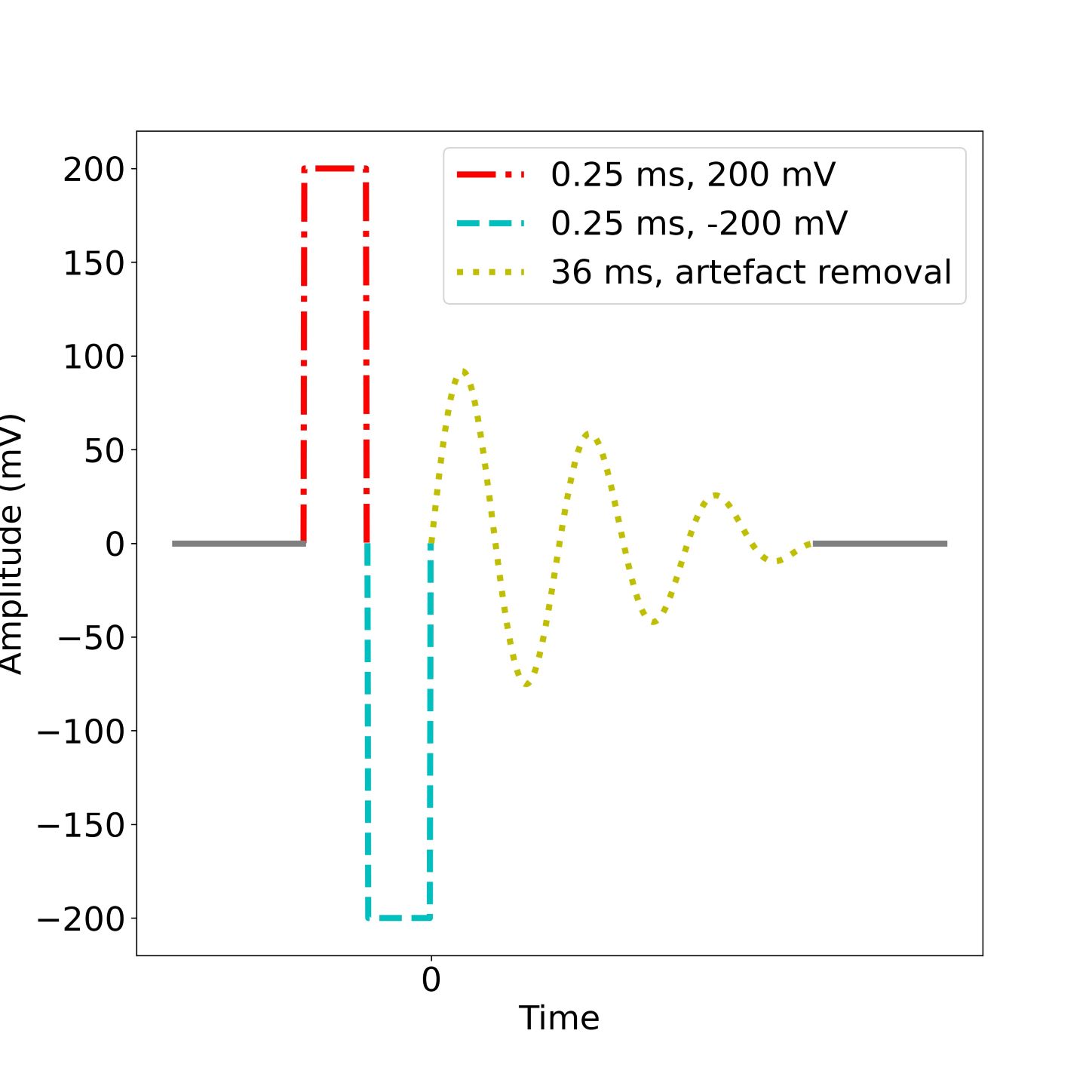}
    \caption{Components of the stimulation pulse. All stimulation pulses were identical and consisted of a bi-polar stimulation lasting 0.5 ms directly followed by a 36 ms artefact removal stimulation. Recorded time of stimulation was set to the end of the bi-polar pulse.}
    \label{fig:stimPulse}
\end{figure}

\begin{figure}
    \includegraphics[width = \linewidth]{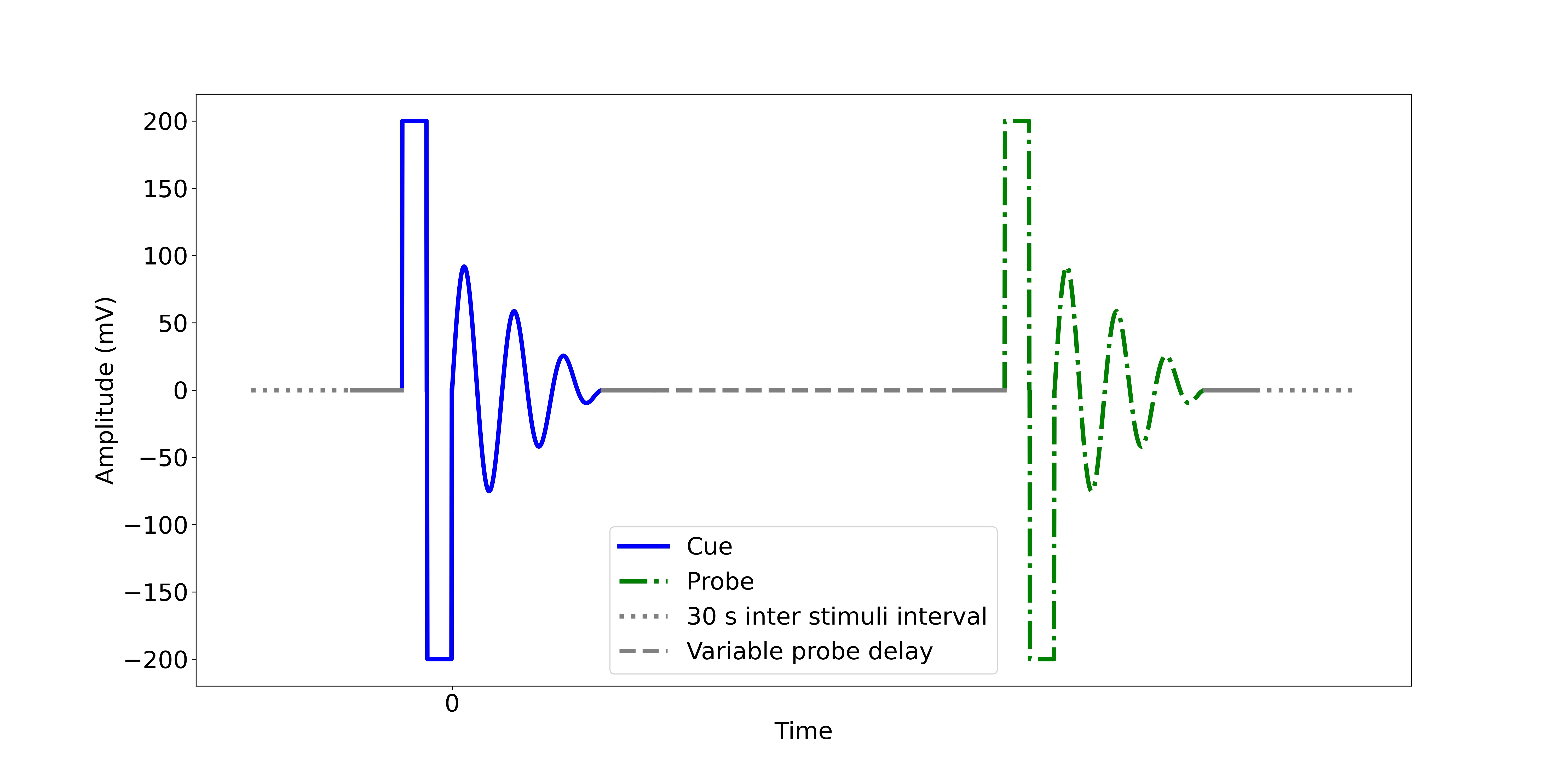}
    \caption{Experimental trial. Each trial consisted of two stimulation pulses input to the network after a 30 seconds rest period. The trial itself consisted of an initial cue stimulation pulse, followed by a probe stimulation pulse at variable delays. In addition a control condition was used where no probe stimulation was given}
    \label{fig:expTrial}
\end{figure}

We conducted two experiments where the timing between two stimulation pulses were varied.  In both experiments the general design of an experimental trial consisted of two identical stimulation pulses, an initial cue stimulation followed by a probe stimulation at variable delays relative to the cue. As a control condition we utilized trials consisting only of a cue stimulation. Details of the stimulation pulse is given in Figure \ref{fig:stimPulse} and an example experimental trial can be seen in Figure \ref{fig:expTrial}.

When referencing the delay between the cue and probe stimulation, i.e. the probe delay, in this text, we refer to the time between the end of electrical activity caused by the artefact removal of the cue and the start of the electrical activity caused by the probe stimulation. Note that given the length of the artifact removal, the minimal distance between two bi-phasic stimulations is 36 ms. The cue stimulation end to probe stimulation start will be given in parenthesis after the probe delay. Both the stimulations in a trial were given to a central electrode at position column 4, row 5 (indexed from 1, from left and top) to maximize our ability to capture the evoked response.

We originally planned to collect data for a unified dataset over two days. Due to limitations of the stimulation software we could not algorithmically create fully randomized condition orders. Instead we opted to use a counterbalancing procedure. 
Each experiment was run over 2 days, with experiment 1 occurring on DIV 24 and 25, and experiment 2 occurring on DIV 27 and 28. During the first day we created a condition sequence starting with a cue-only control trial which was followed by a single trial from each condition, in order of increasing probe delay. Each trial was preceded by a 30 s rest period as indicated previously and the condition sequence was repeated 60 times. On the second day the condition sequence was reversed such that the first trial in the sequence was from the condition of largest probe delay and was followed by probe delays in decreasing order, ending with the cue-only control condition. With 60 repetitions each day, this resulted in 120 trials for each condition (including the cue-only control condition).

\subsubsection{Experiment 1}
\label{sec:Exp1}
In experiment 1 we targeted the specific phases of neural responses identified by \cite{wagenaar2004effective} and \cite{wagenaar2006extremely}: 

\begin{itemize}
    \item 0 (36) ms - early post synaptic phase (5 - 50 ms)
    \item 100 (136) ms - start of network wide barrages (0 - 100+ ms)
    \item 1000 (1036) ms - upper mean length of culture wide barrages
    \item 7000 (7036) ms - after networks can produce culture wide barrages post stimulation (5+ s)
\end{itemize}

\subsubsection{Experiment 2}
\label{sec:Exp2}

In experiment 2 we did not target specific phases but instead linearly increased the probe delay in steps of 200 ms starting at 200 ms up to 800 ms to target the period with active cue evoked activity. This period was also identified as providing high classification accuracy, we thus were interested in exploring the pattern of separability and acuity during this period: 

\begin{itemize}
    \item 200 (236) ms
    \item 400 (436) ms
    \item 600 (636) ms
    \item 800 (836) ms
\end{itemize}

\newpage
\section{Analysis and Results}
\label{sec:AnRes}
As we investigate 5 different features related to our encoding method and computing system we divide the Analysis and Results section into 5 sub-sections related to each topic:

\begin{enumerate}
    \item \hyperref[sec:NetStab]{Network Stability}
    \item \hyperref[sec:DecReadInfCode]{Decoder-Readout and Information Encoding}
    \item \hyperref[sec:UpMem]{Upper Separablity Bound}
    \item \hyperref[sec:LowSepBound]{Lower Separability Bound}
    \item \hyperref[sec:PPAcu]{Probe - Probe Acuity}
\end{enumerate}

In the two first sub-sections  we describe two research questions that have downstream effects on the three latter. 
First we assess the stability of the networks over their development and from day one and two of each experiment. Next, because the optimal decoder-readout parameters may differ for each research question, we describe the parameter options and their relevance.
These sub-sections are followed by the three sub-sections related to our main research questions, the upper and lower separability bounds and the acuity of the network response to different stimulation timings. Each are divided into a short introduction, where we elaborate the specific goal and justification for the analysis, a methods section, where we describe the classification task, a results section where we present the results for the research question, the optimal encoder settings and related neural spike code features and finally a brief discussion of the results. A summary discussion of the overall results follows in the general discussion section.

\subsection{Network Stability}
\label{sec:NetStab}
For in-vitro neural networks to be usable for practical applications they must display relatively consistent behavior over time. It would be rather impractical if a self driving car recognized buildings as buildings one day and buildings as highways the next. Therefore, it is important to explore how the networks' behavior change over time. Since we ran each experiment over the course of two days, instability could specifically interfere with the decoder's ability to learn from the responses of the network if they are inconsistent between the two days. Multiple factors could cause such instabilities. In-vitro neural networks goes through different developmental stages, starting with a lack of spiking, moving onto bursting and then often developing into critical dynamics, measured through avalanche distribution \cite{tetzlaff2010self} meaning that the network response could potentially change drastically over time. Furthermore, the networks are only fed every few days and do not have the benefit of continuous metabolic support that would normally ensure homeostatic conditions. 

\subsubsection{Methods}
\label{sec:NethStabMeth}
We extracted overall bursting and spiking behavior during a 1 hour daily baseline recordings and plotted the mean number of burst and spikes during these periods for the extent of the network's lifetime. 
Next we plotted the responses to stimulation between days for each experiment based on 1024 ms epochs locked to the offset of the probe/control stimulation. This was done by both averaging over electrodes, time bins and trials (Figure \ref{fig:MeanProbeResp}) and only averaging over time bins and trials (Figure \ref{fig:MeanFeature1000}). Lastly we plotted the epoched data at single trial level to visually inspect the responses (available in the digital repository: \url{https://osf.io/yzf47/})
We also computed the total number of spikes evoked by each experimental condition for each day. We tested the data for normality using Shapiro 
-Wilk tests and homogeneity of variance using Levene's test. Based on the results of these tests we opted for non-parametric alternatives to student's t-test. First inspecting boxplots of the data, they appeared not to show similarity in distribution, we therefor used Mann-Withney U tests, interpreting the results as tests for differences in distribution. Finally we performed Welch's t-tests with 10 000 permutations to test for differences in the evoked response. All tests where performed using scipy.stats (version: 1.9.3). P-values from the Welch's t-test are presented in table \ref{tab:Table1}, for  extended results see the digital repository at \url{https://osf.io/7p4rh}.
Finally, for the upper and lower separability bound and separability acuity we tested each day separately to visualize the change in performance.
The two first points are plotted below, while the final is shown within their respective sections.

\subsubsection{Results}
\label{sec:NetStabRes}
\begin{figure}[!htb]
  \centering
  \subfloat{
    \includegraphics[width = 0.8\linewidth]{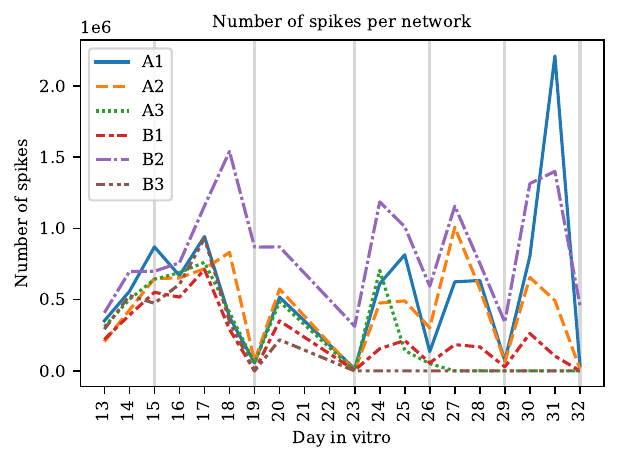}
    \label{fig:MeanSpike}
  }
  \vfill
  \subfloat{
    \includegraphics[width = 0.8\linewidth]{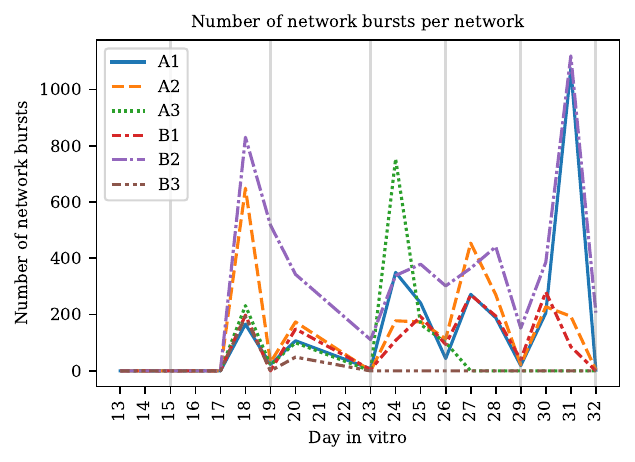}
    \label{fig:MeanBurst}
  }
  \caption{Sum spikes (top) and burst (bottom) during the 1 hour baseline recording. Feeding days are marked as vertical lines }
  \label{fig:MeanSpikeAndBurst}
\end{figure}

Mean spiking activity increased gradually until around DIV 17 as seen in Figure \ref{fig:MeanSpike}, after which activity decreased before fluctuating in tandem with the days in which the nutrient medium was exchanged (indicated as vertical lines in Figures \ref{fig:MeanSpike} and \ref{fig:MeanBurst}). Networks A3 (green ... line) and B3 (brown -..- line) also appeared to see a rapid activity decline starting at DIV 23 for B3 and DIV 27 for A3.
Similar behavior is seen for bursting behavior in Figure \ref{fig:MeanBurst} where bursts first arise at DIV 18 and then fluctuates with medium exchange days while A3 and B3 declined rapidly.

\begin{figure}[!htb]
    \centering{}
    \includegraphics[width =1\linewidth]{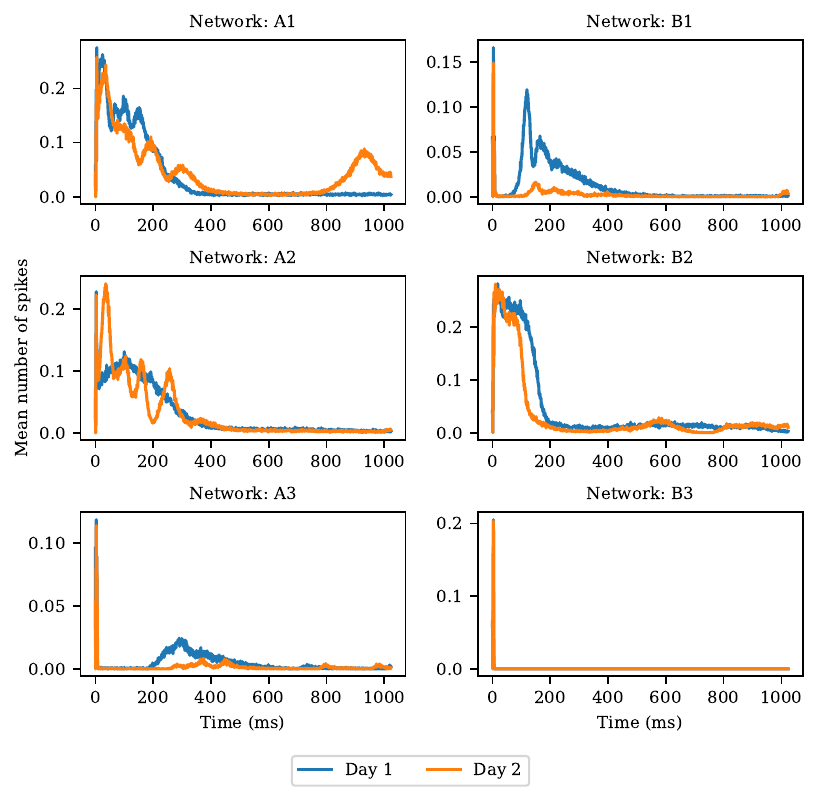}
    \caption{Mean cue evoked response, averaged over electrodes at 1 ms time bins, across day 1 and 2 of experiment 1, control condition. Plots for additional conditions can be found in the digital repository: }
    \label{fig:MeanProbeResp}
\end{figure}

Probe evoked responses also differed between days. In Figure \ref{fig:MeanProbeResp} we show the mean number of spikes averaged over electrodes and time bins (1 ms) during an epoch after the control conditions in experiment 1, which can be seen to vary considerably between days. We found that for multiple conditions and networks the number of evoked spikes within a 1024 ms epoch after the stimulation was significantly different based on Welch's t-test with 10 000 permutations (see table \ref{table:EvokedDiff}) 

\begin{table}[!htb]
\caption{p-value statistics for the day to day difference between probe evoked number of spikes for each well and condition for experiment 1. Values are rounded to 4 decimal points.}
\label{tab:Table1}
\begin{center}
\begin{threeparttable}
\begin{tabular}[width = 0.8\linewidth]{lcccccc}
\toprule
 & \text{A1} & \text{A2} & \text{A3} & \text{B1} & \text{B2} & \text{B3} \\
\midrule
\text{control} & 0.0003** & 0.6705 & 0.0004** & 0.0001** & 0.0001** & 0.0096* \\
\text{0 ms} & 0.0001** & 0.1542 & 0.0001** & 0.0001** & 0.0001** & 0.0371* \\
\text{100 ms} & 0.0001** & 0.3149 & 0.0001** & 0.0001** & 0.0001** & 0.0066* \\ \\
\text{1000 ms} & 0.0001** & 0.0001** & 0.0272* & 0.0446 & 0.0001** & 0.0204* \\
\text{7000 ms} & 0.0064* & 0.0184* & 0.0001** & 0.0001** & 0.0001** & 0.1443 \\
\bottomrule
\end{tabular}
\begin{tablenotes}
\small
\item Note: * indicates p $<$ 0.05, ** indicates bonferroni corrected p $<$ 0.00167
\end{tablenotes}
\end{threeparttable}
\end{center}
\label{table:EvokedDiff}
\end{table}

\begin{figure}[!htb]
    \centering{}
    \includegraphics[width = \linewidth]{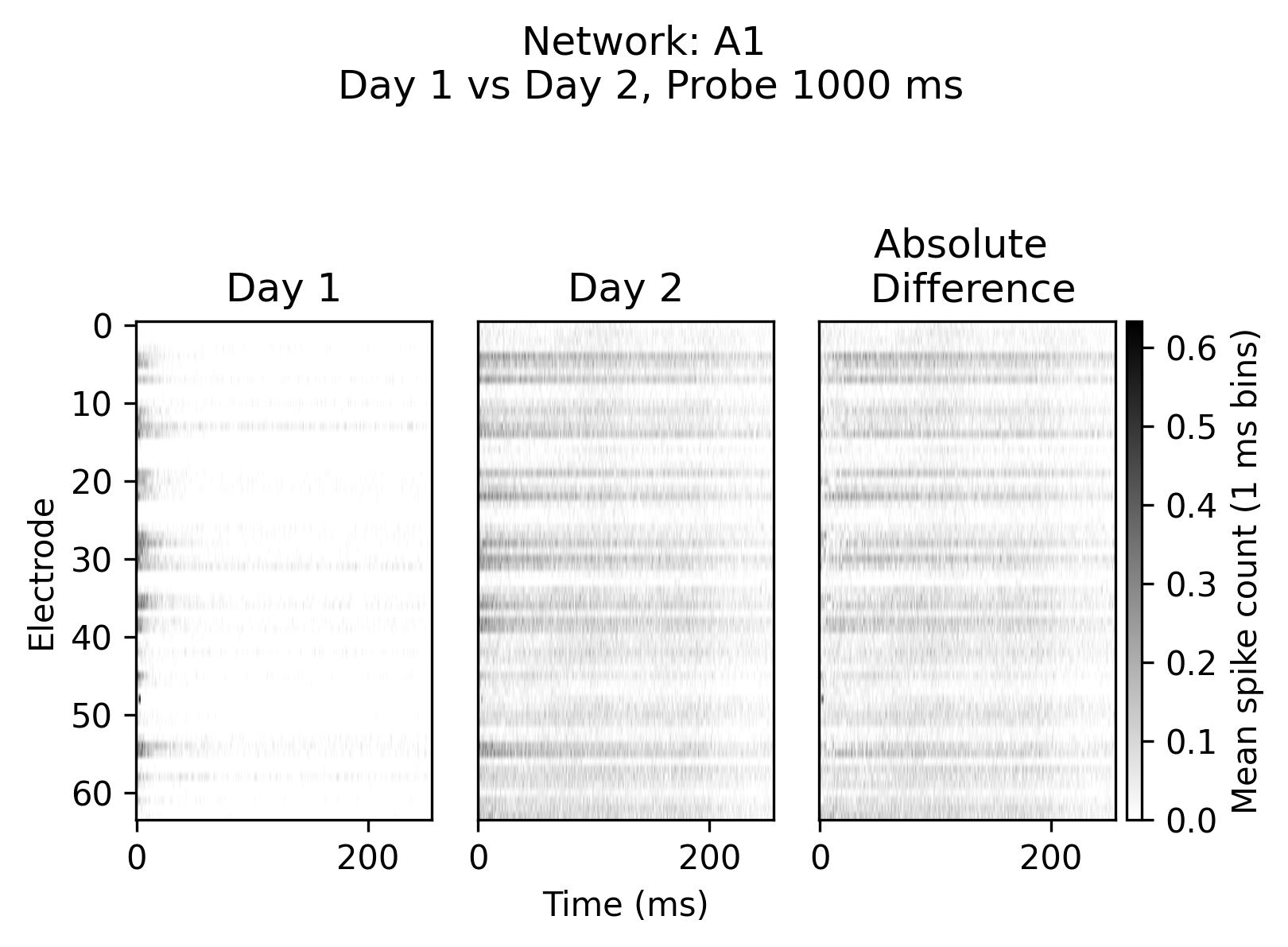}
    \caption{Mean features for 1000 ms probe delay averaged over trials at 1 ms time-bins. The leftmost image shows the evoked response during day 1, the middle shows the response to day 2 and the rightmost panel shows the absolute difference between the two responses. Epoch size was 256 ms and time bin size = 1}
    \label{fig:MeanFeature1000}
\end{figure}

To further visualize the difference between days we also create mean evoked responses at the electrode level. In Figure \ref{fig:MeanFeature1000} we show the mean responses to day one and day two of experiment 1 on the left and center and the difference between the two on the right.

\subsubsection{Discussion}
\label{sec:NetStabDis}
The fluctuating activity level, both in the form of bursts and in the form of spiking activity, and its concurrence with feeding days may indicate a core issue with the disembodied nature of in-vitro neuronal networks. Neurons in the brain receive increased blood flow carrying nutrients and oxygen when their activity increases \cite{magistretti2015cellular}. Conventional in-vitro neural networks rely on the availability of nutrients through cell culture media for their energy demands. Replenishing such media may affect neuronal activity, both in  terms of adjusting the availability of nutrients relative to energy demands within the network, but also as a result of potential mechanical perturbation to the cells during media change and aseptic handling \cite{heo2016dependence}. This may account for the differences in network dynamics observed in the period directly following media replenishment but may also have affected the networks´' ability to produce consistent longer term activity, which in turn becomes a challenge  due to the change in evoked responses. A classifier trained on responses at a given time window after media change may thus not be able to classify responses at any other time-window after media change due to the underlying change in network dynamics. This potential issue is seen clearly in the mean features of Figure \ref{fig:MeanFeature1000}. However, it may be possible that consistent activity patterns persists even though overall activity level changes. This warrants the comparison of actual performance between experimental days which is tackled in the following sections. Furthermore it indicates that we may not combine data from the two days the experiments were ran over as originally planned.

\subsection{Decoder-Readout and Information Encoding}
\label{sec:DecReadInfCode}
As mentioned in the introduction and experimental method sections, \cite{wagenaar2004effective} found different phases of spike responses to electrical stimulation. The earliest phase consisted of temporally consistent spikes, while the consecutive phases consisted of high spike counts but with lower temporal consistency, suggesting two different information codes, namely spike-time and rate.

This is important for the choice of readout parameters.
If the network uses rate based encoding we can capture this information by binning spikes over large time intervals, i.e using large time bins. On the other hand, if the network uses spike time encoding, large time bins would delete the specific temporal information of the spikes. This would mean that we would be unable to use this information for classification. However, if we use small time bin sizes we would be able to capture spike time encoding, but may be unable to capture rate based encoding.

The networks may also use both types at the same time as these coding methods can easily be multiplexed i.e both the rate increases and the specific spike times are consistent but different. It is also conceivable that the coding method may change over time given that the evoked response observed in \cite{wagenaar2004effective, baljon2009interaction} had distinct phases with more or less consistent spike patterns.

\subsubsection{Methods}
\label{sec:DecReadInfoCodeMeth}

\begin{figure}[!htb]
    \centering{}
    \includegraphics[width=0.7\linewidth]{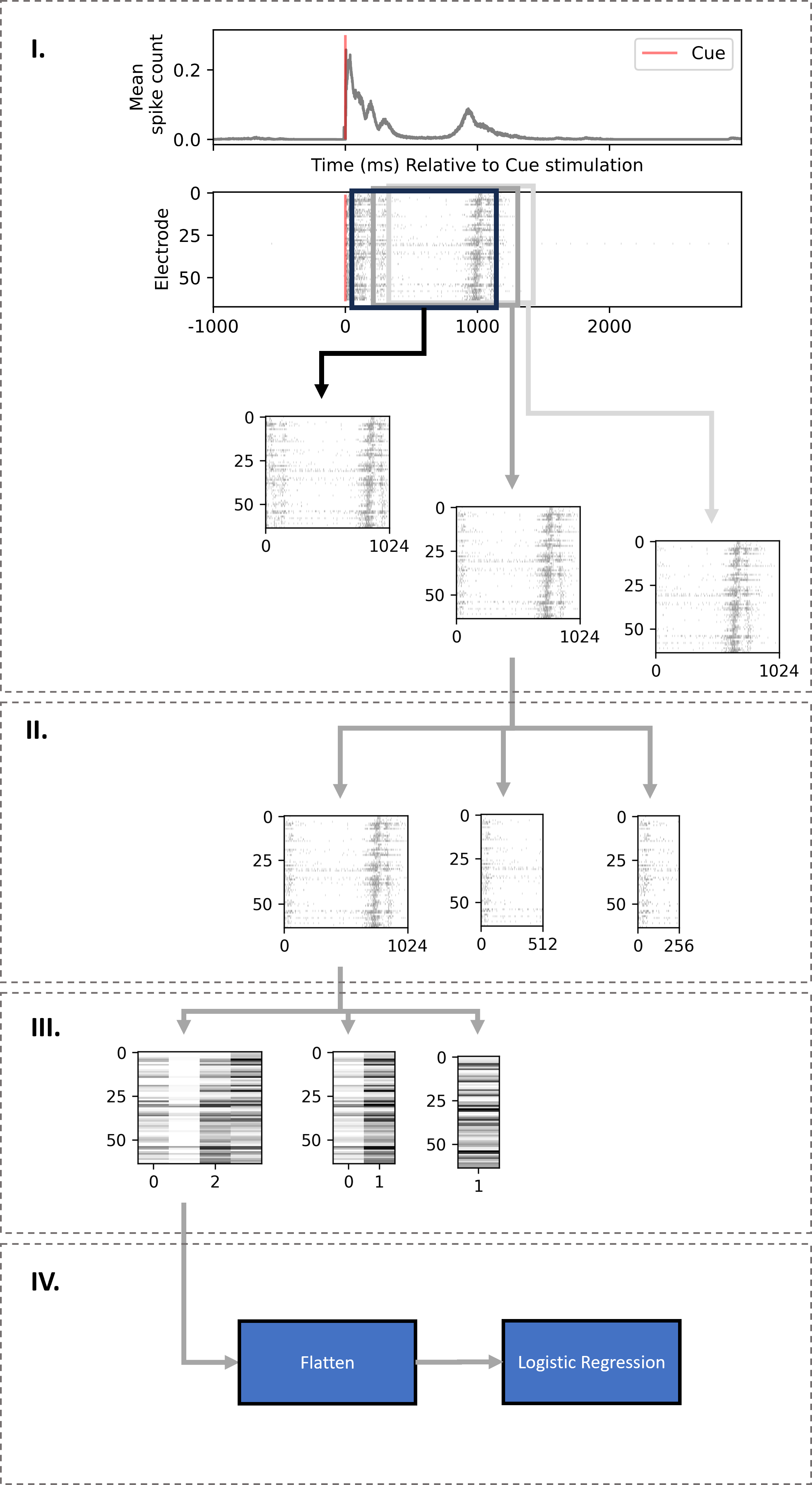}
    \caption{Grid search over the Decoder-Readout parameters. $\mathbf{I.}$ Epochs of 1024 ms are extracted with different offsets relative to the target stimulation. $\mathbf{II.}$ Epochs length is varied by reducing the length of the original extracted epoch. $\mathbf{III.}$ Spikes are time binned into different time bin sizes. $\mathbf{IV.}$ Time binned epochs are flattened before being compiled to training and test data and passed to the logistic regression model}
    \label{fig:PreProssPipeLine}
\end{figure}

To find the optimal readout parameter settings for our different tasks we performed a grid search through a subset of the parameter space. We varied the epoch length, time bins size and epoch offset, summarized in Figure \ref{fig:PreProssPipeLine}.
Epoch length (time window relative to stimulation time, during which we extract spikes) was varied from 1 ms to 1024 ms in powers of 2, similarly we varied the number of time bins from 1 ms up to the epoch size in powers of 2. 
The epoch offset was varied from 1 ms up to 2048 ms in steps of 1 ms.
By varying the epoch length we explore how much of the evoked response is needed to classify the responses and allowed for multiple variations of time bin size. The time bins were varied to identify whether the current epoch carried information in spike times or rate. Importantly, varying the epoch length and time bin size in powers of 2 ensures that any time bin of a larger size can be created from time bins of any smaller size without any smaller time bin being split. Building features using this method ensures that we do not create any new features by splitting smaller features, which could potentially cause the comparison between larger and smaller time bins to become inaccurate as we would be creating new information instead of just destroying it as is done when two time bins are combined. 
The epoch offsets was used to identify how the encoded features changed over the temporal extent of the evoked responses. This also allows us to see how long the information is available in the response.
As the optimal settings may vary for each of our analysis, results related to these methods are described within the upper, lower and acuity separability sections.

\newpage

\subsection{Upper Separability Bound}
\label{sec:UpMem}

For our main research questions we first attempt to find the upper bound of the linearly separable space of stimulation timing. We hypothesize that this bound may be defined by the memory length of the networks. Importantly, this may include both "visible" memories, i.e spiking activity visible through the recording electrode, and "hidden" memories i.e. memories not visible through the electrode. Examples of hidden memories may include short term synaptic depression, facilitation or neural exhaustion present after evoked activity has settled. Such memories have been identified in \cite{dranias2013short}.

If a memory of an initial stimulation (the cue) persists in the network, a secondary stimulation (the probe) may interact with this memory. This interaction may produce a response that is linearly separable from responses evoked by a single stimulation given in isolation. Conversely, if the memory has faded completely, no interaction can take place and the response should be inseparable from the response produced by an isolated stimulation. 

\subsubsection{Methods}
\label{sec:UpMemMeth}
\begin{figure}[!htb]
    \centering{}
    \includegraphics[width=1\linewidth]{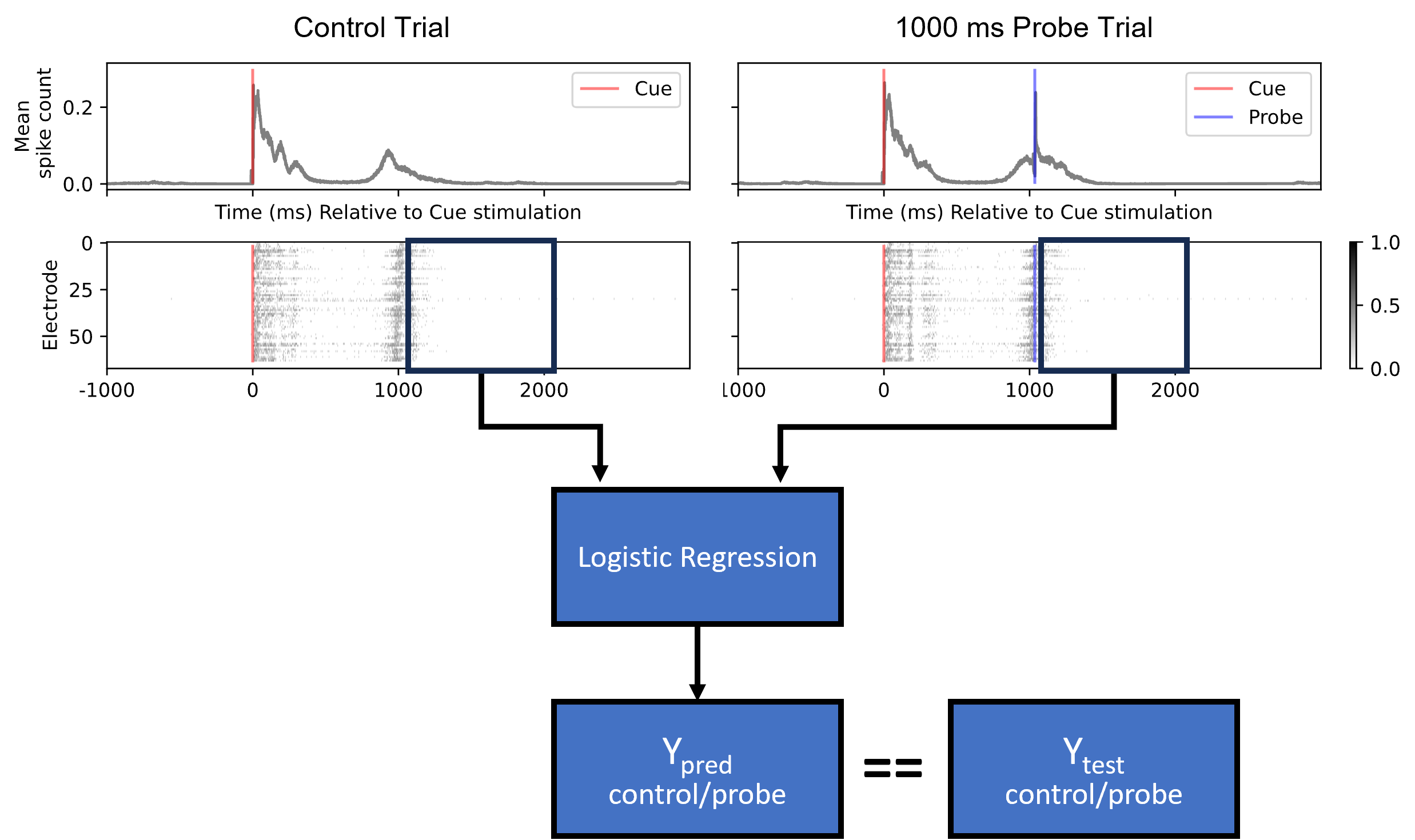}
    \caption{Epoch selection for the binary classification task assessing the upper separability of stimulation timing encoding. The top two plots show spike counts per ms averaged over electrodes and trials for the control and 1000 ms probe delay condition. An example trial from each condition is given below and the epoch window for a 1024 ms epoch is indicated as a black square outline. Epoch extraction and prepossessing is conducted according to \ref{sec:DecReadInfoCodeMeth} before trials are passed to the logistic regression model. Predictions $y_{pred}$ are output as labels identifying the intput as either originating from a control cue stimulation or experimental probe stimulation.}
    \label{fig:UpSepBoundPipeline}
\end{figure}

Epochs offset relative to probe from the experimental trials and from the cue in the control trials, were extracted from Experiment 1. We created binary classification tasks between each experimental condition and the control condition. Epoch positions are visualized in Figure \ref{fig:UpSepBoundPipeline}.

The classifier was thus tasked with classifying whether a given trial belonged to the cue of the control trial or to the probe of an experimental trial. With 4 experimental conditions this resulted in 4 separate binary classification tasks.

\subsubsection{Results}
\label{sec:UpMemRes}

\begin{figure}[!htb]
    \centering{}
    \includegraphics[width = 0.8\linewidth]{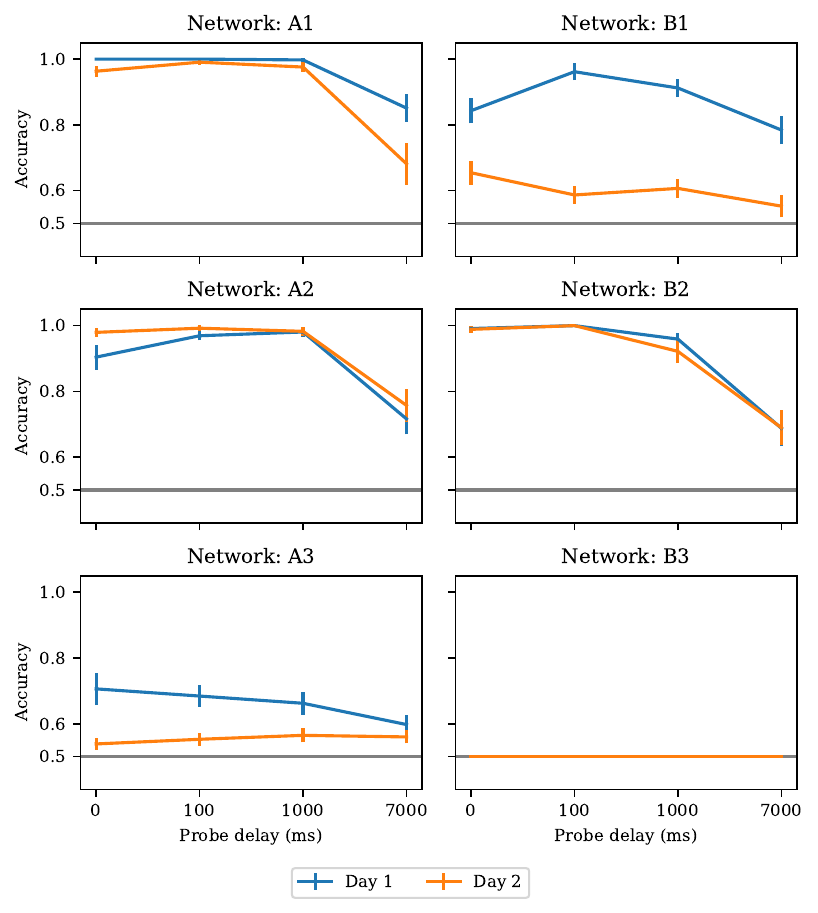}
    \caption{Mean accuracy for the highest performing combination of epoch size, time-bin size and epoch offset for the binary classification of probe vs cue stimulations. Results from day 1 is shown in blue and day 2 in orange. Standard deviations are given as vertical bars. }
    \label{fig:upperSep}
\end{figure}

A summary of the highest performing decoder-readout setting results can be seen in Figure \ref{fig:upperSep}, and the specific combination of offset, epoch and time bin size for each condition can be found in Table \ref{ap:upSepD1} and \ref{ap:upSepD2} in the appendix. We observe mixed performance across networks, days and conditions. Three networks showed high performance for the 0, 100 and 1000 ms probe conditions but decreased performance for the 7000 ms condition. However, some still performed above chance level. Network A1 performed with 85\% accuracy for the 7000 ms condition on one day. A3 and B3 performed poorly overall while B1 performed relatively good on day 1 and poorly on day 2.

\begin{figure}[!htb]
    \centering{}
    \includegraphics[width = 1\linewidth]{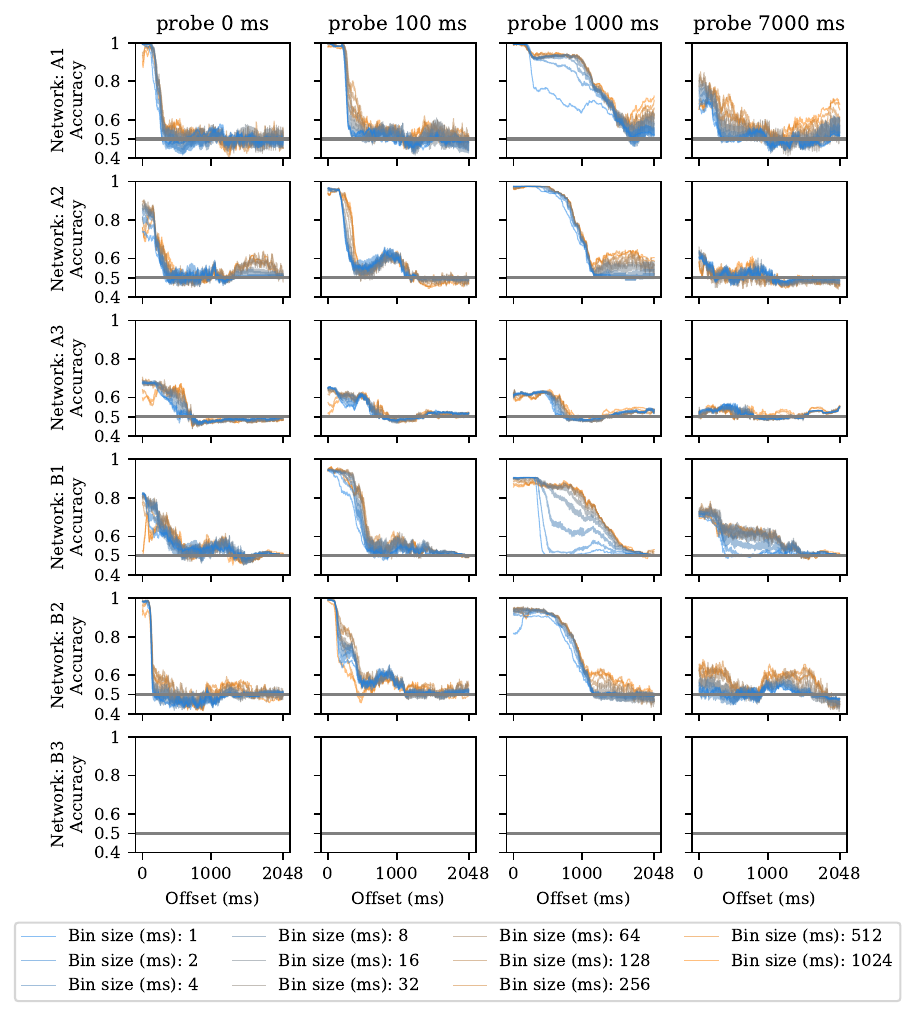}
    \caption{Mean accuracy from binary classification between probe vs cue stimulations shown for epoch size = 1024 and all time bins and offsets. Results are given for day 1, with day 2 available in the digital repository }
    \label{fig:upperSepOffsets}
\end{figure}

\begin{figure}[!htb]
    \centering{}
    \includegraphics[width = 1\linewidth]{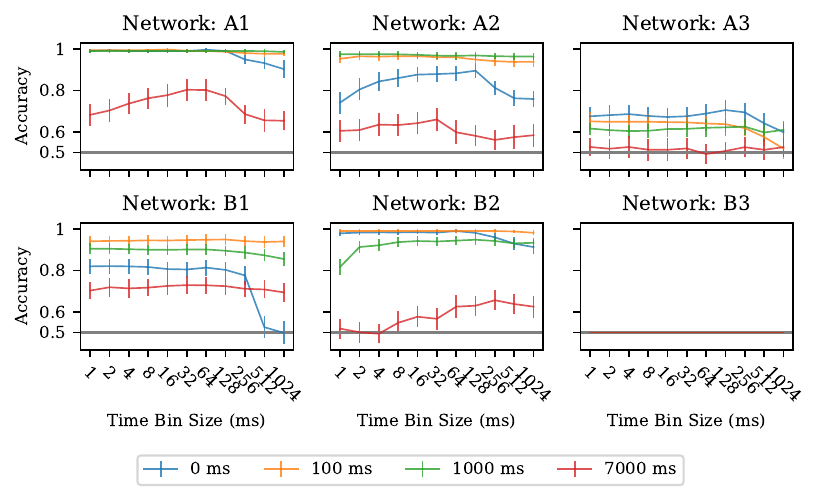}
    \caption{Mean accuracy of the binary classification between cue and probe evoked responses shown for offset = 0 and epoch size = 1024 to visualize the effect of time bin size. Probe delay is given in legend.}
    \label{fig:TB_effect_upper}
\end{figure}

Exploring the classification accuracy over offsets revealed that the accuracy of the classifier stayed relatively stable for a short time before falling off, the length of this stability was however varied. This can be seen in Figure \ref{fig:upperSepOffsets} were we show the classifiers' accuracy for all time bins for epoch size = 1024 ms. The different time bin sizes are shown in gradient colors from blue (1 ms time bin) to orange (1024 ms time bin). We can here see different effects depending on the network and condition with small time bins decreasing in accuracy more rapidly than larger for the 1000 ms probe delay condition in A1 and B1. Focusing on the non-offset epoch as visualized in Figure \ref{fig:TB_effect_upper}, we see the different effects of time bin size on the different networks and conditions. In some cases the effect is minimal as is the case for the 100 and 1000 ms conditions in network A1, A2, B1 and B2, while the 0 and 7000 ms conditions showed more instability.

\subsubsection{Discussion}
\label{sec:UpMemDis}
Results show that performance is varied but memory lengths can exceed 7 seconds. This suggests that we may be able to input data through stimulation timing over a very wide range of stimulation delays, although at the risk of lower accuracy. Exactly how long these delays can be must be established on a network to network basis as we see varied performance across networks. Furthermore, the day to day instability in the neural response may be problematic as this may interfere with the generalizability of the classifier over time. 

\subsection{Lower Separability Bound}
\label{sec:LowSepBound}

Our next goal was to identify the lower bound of the linearly separable stimulation timing space and to ensure that the probe stimulation did affect the trajectory of the networks. 

A single neuron will experience a refractory period after producing a spike during which new input will struggle to elicit a new spike \cite{vardi2021significant}. Thus, single neurons have a lower limit to the separability of its input space. 
At the network level, low frequency stimulation often elicit a population burst \cite{wagenaar2005controlling}, this could cause the network to be saturated by the activity caused by the cue stimulation such that the probe is unable to produce any network activity of its own. 
This is important for assessing the minimal timing difference one can use to encode information. If the probe is not able to affect the network below some minimal timing difference this sets a limit for whether a third stimulation could interact with the probe and cue evoked activity which will limit the information encoding rate we can use.

\subsubsection{Method}
\label{sec:LowSepBoundMeth}
The method described in the previous section can only be used to detect when the effect of a previous stimulation is gone from the network and cannot be used to directly explore interactions between inputs because the classifier may use the position in the cue evoked response for classification even if the probe had no effect on the network activity. 

Instead of comparing probe locked activity to cue locked activity, we extract epochs from the cue response locked to the time of the probe. This way the classifier cannot "cheat" by using the "time since stimulation" computation embedded in the cue response as it is identical if the probe had no affect on the network.

To ensure that we capture the exact same epoch relative to the cue stimulations we first extracted cue locked responses for all conditions and recalculated spike times relative to cue stimulation time.
For the experimental trials, which contain a probe stimulation, we then extracted a new epoch from the cue locked epoch, starting at the time of the probe + 36 ms (to avoid capturing artefacts caused by the stimulation and artefact removal). 

For each experimental condition we also extracted epochs from the cue-only control condition at the time corresponding to each probe + 36 ms in the experimental conditions. 

Other than the training data the training task was identical to the one described for the upper separability bound.
\begin{figure}[!htb]
    \centering{}
    \includegraphics[width=\linewidth]{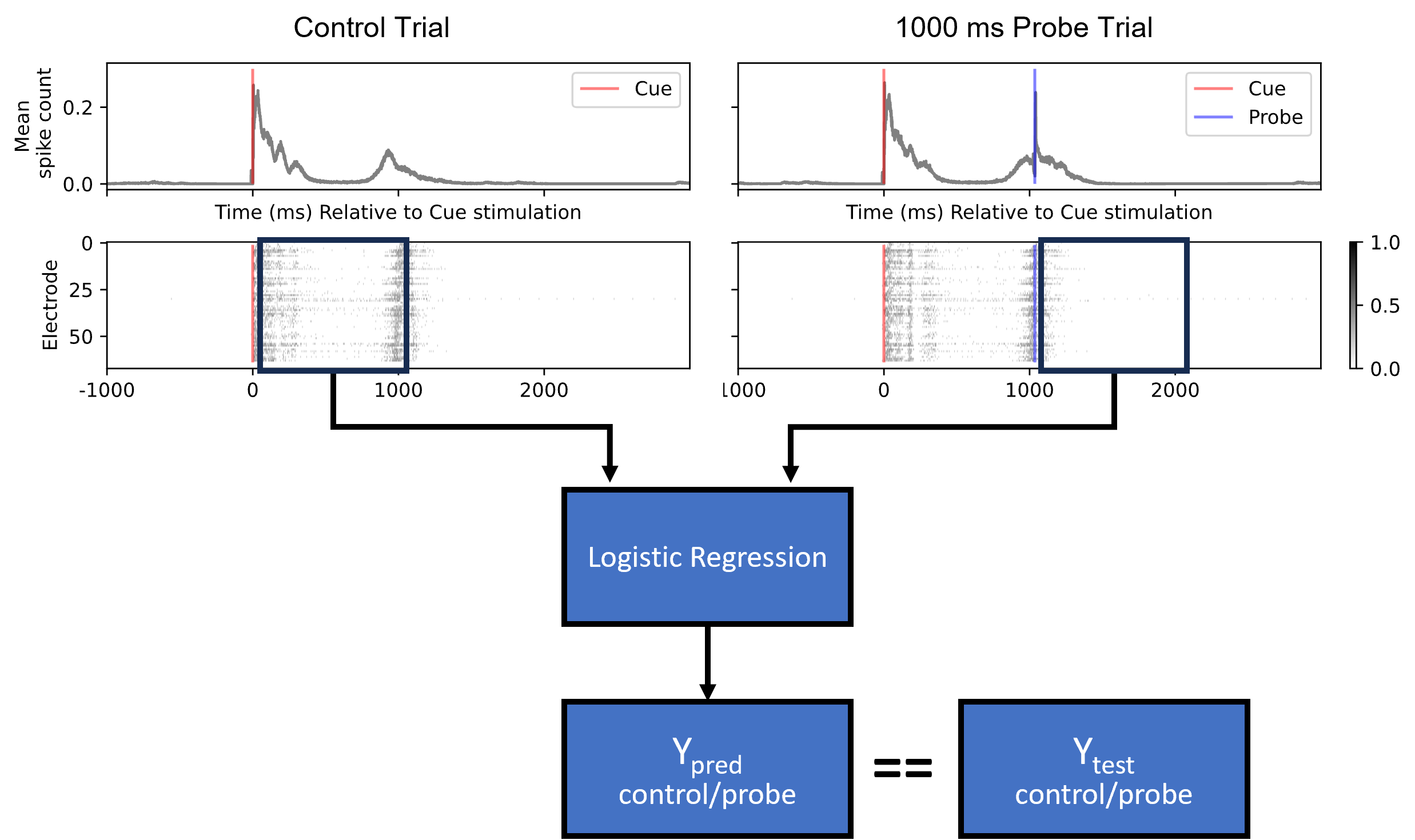}
    \caption{Epoch selection for the binary classification task assessing the lower separability of stimulation timing encoding. The top two plots show spike counts per ms averaged over electrodes and trials for the control and 1000 ms probe delay condition. An example trial from each condition is given below and the epoch window for a 1024 ms epoch is indicated as a black square outline. The epoch from the control condition is here offset from the cue stimulation so that it matches the position of the epoch from experimental condition. Epoch extraction and prepossessing is conducted according to \ref{sec:DecReadInfoCodeMeth} before trials are passed to the logistic regression model. Predictions $y_{pred}$ are output as labels identifying the input as either originating from a control cue stimulation or experimental probe stimulation.}
    \label{fig:LowerSepBoundPipeline}
\end{figure}

\subsubsection{Results}
\label{sec:LowSepBoundRes}

Max performance was in some cases lowest at the smaller probe delays (A1, B2, both days and A2 day 2), B1 showed a curved pattern with lower accuracy at 100 and 1000 ms probed delays at both days. A2 at day 1 showed decreasing accuracy from 0 ms probe delay but peaked at 7000 ms probe delay as can be seen in Figure \ref{fig:sepLower} and Table \ref{ap:lowSepD1} and \ref{ap:lowSepD2} in the appendix. A3 and B3 again had poor accuracy. 
At the smallest probe delay of 0 ms, A2 had the highest accuracy at 75 \% on day 1, this was however decreased to 64 \% for the 100 ms probe. B2 was the best performing network at 100 ms probe delay with an accuracy of 79 \%.
\begin{figure}[!htb]
    \centering{}
    \includegraphics[width = 0.8\linewidth]{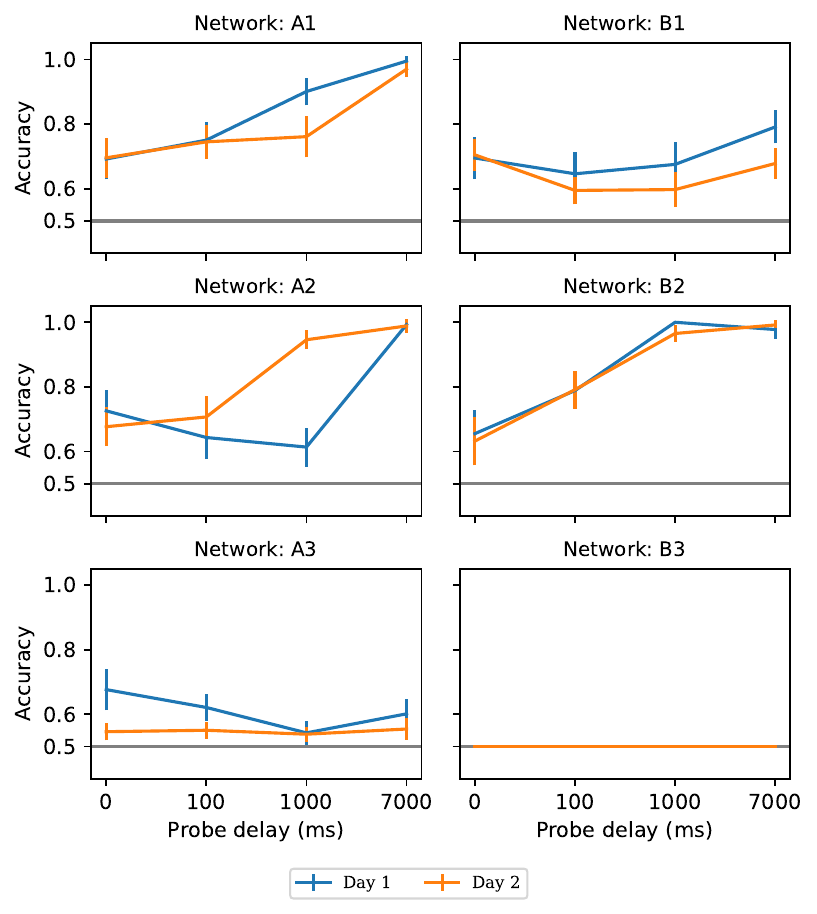}
    \caption{Mean accuracy for the highest performing combination of epoch size, time-bin
size and epoch offset for the binary classification of control vs experimental conditions, extracted at identical times relative to cue stimulation. Results
from day 1 is shown in blue and day 2 in orange. Standard deviations are given as vertical
bars  }
    \label{fig:sepLower}
\end{figure}

\begin{figure}[!htb]
    \centering{}
    \includegraphics[width = 1\linewidth]{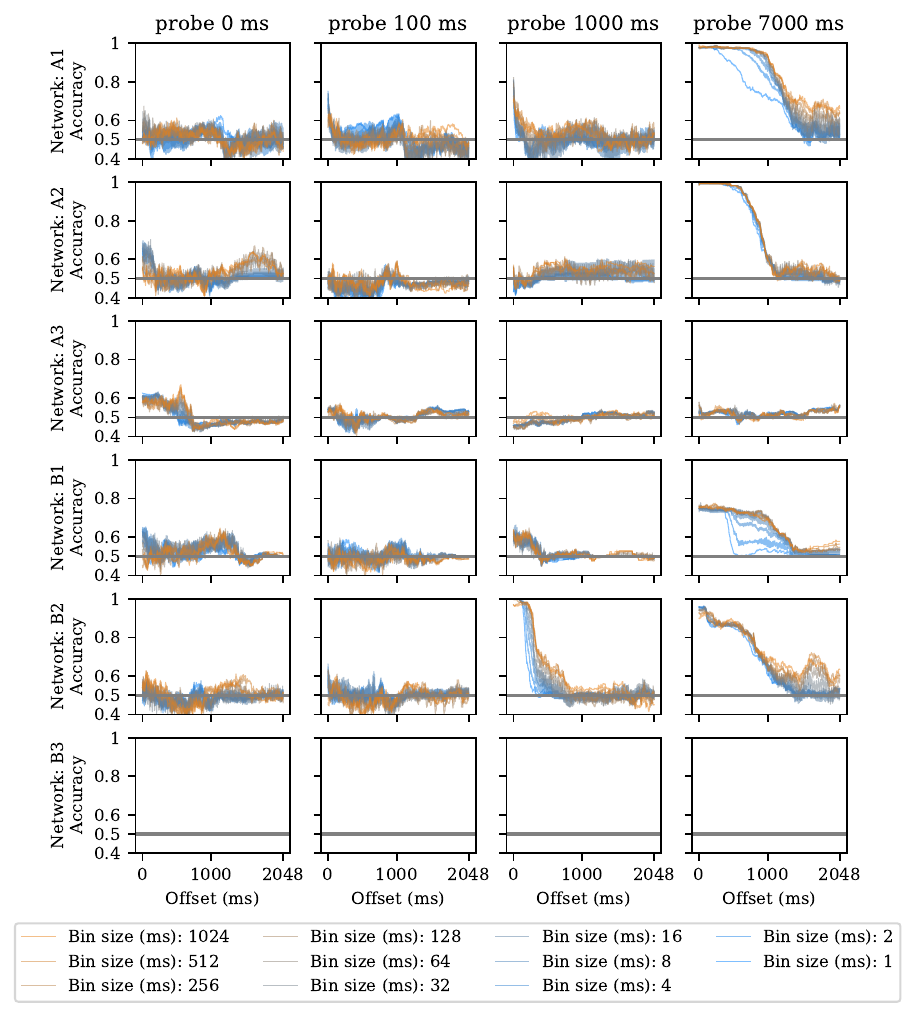}
    \caption{Mean accuracy from binary classification of control vs experimental conditions, extracted at identical times relative to cue stimulation
shown for epoch size = 1024 and all time bins and offsets. Results are given for day 1,
with day 2 available in the digital repository
}
    \label{fig:lowLepOverOffset}
\end{figure}

Visualizing the accuracy over epoch offsets for an epoch size of 1024 ms in Figure \ref{fig:lowLepOverOffset}, we can see how the accuracy of the classifier changes over the evoked neural responses.

In some cases, in particular for A1, higher accuracy is concentrated relatively early with small offsets and is not sustained for extended periods. Generally however, accuracy is poor for most offsets. The 7000 ms probe condition (and 1000 ms for B2) does show high accuracy and for multiple networks this is sustained for some time before falling off. Here we can also see that the sustain in accuracy is better with larger time bins, with the smaller time bins dropping in accuracy faster than the larger as is the case for A1 and B1.

\begin{figure}[!htb]
    \centering{}
    \includegraphics[width = 1\linewidth]{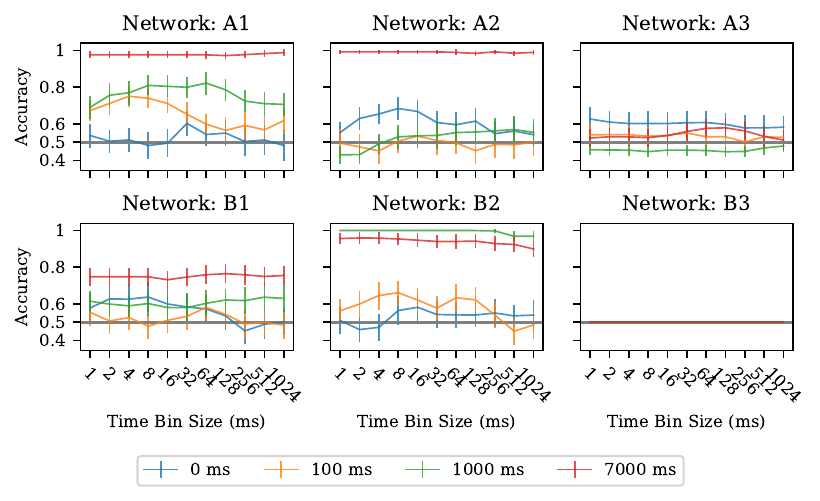}
    \caption{Mean accuracy of the binary classification of control vs experimental conditions, extracted at identical times relative to cue stimulation, shown for offset = 0 and epoch size = 1024 to visualize the effect of time bin size. Probe delay is given in legend.}
    \label{fig:BinSize_lower}
\end{figure}

While classification accuracy in some cases decreased as a function of decreasing time-bin size at later offsets, no particular pattern was found directly after stimulation. In Figure \ref{fig:BinSize_lower} we visualize the drop in accuracy for the first epoch offset (offset = 0 ms) for epoch size = 1024 ms where this can be seen.

\subsubsection{Discussion}
\label{sec:LowSepBoundDis}
The lower accuracy at smaller probe delays may be problematic as it indicates that the probe stimulation did not have sufficient effect on the network activity.
This will make it difficult to compound multiple inputs to produce unique outputs, at least if they are initiated to a network under the same conditions as ours. Furthermore, the effect that was seen, affected the network activity differently at the scale of spike timing and rate. This may indicate that an optimal readout should include multiscale readout time-bins to capture both high accuracy spike time codes and longer memory rate codes.

\subsection{Probe - Probe Acuity}
\label{sec:PPAcu}
While probe responses may be separable from cue responses, two probes
given at different delays may not be separable. Meaning that the probe
response does not encode a specific temporal delay but rather a non-specific
delay. E.g. a specific delay encoding would tell us that a cue-probe delay was of t ms, while a non-specific delay may encode that a cue-probe delay was of t +/- x ms. We thus need to explore the probe-probe separability in addition to
cue-probe separability. In other words the acuity of the network response.

\subsubsection{Methods}
\label{sec:PPAcuMeth}
\begin{figure}[!htb]
    \centering{}
    \includegraphics[width=\linewidth]{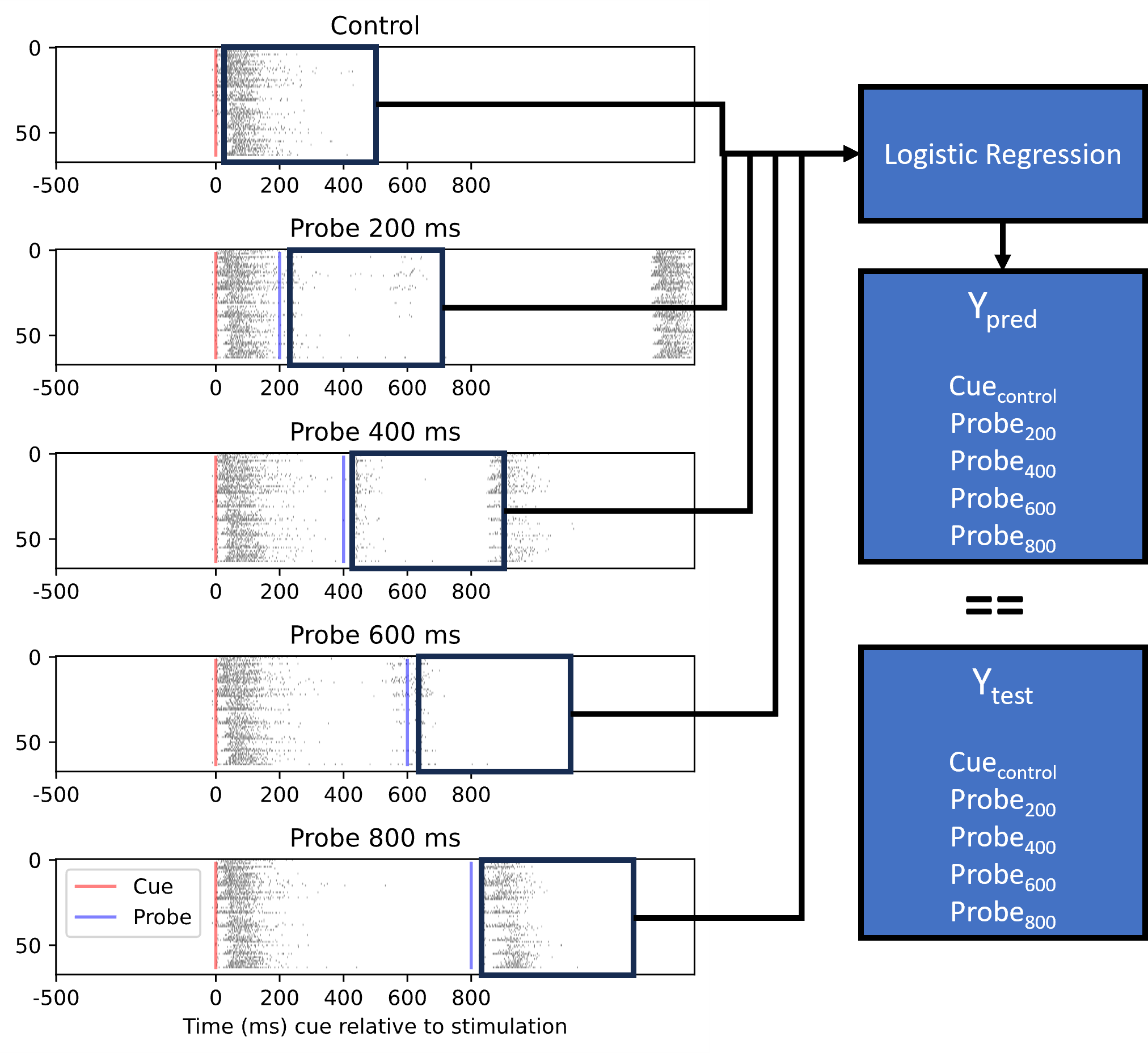}
    \caption{Probe - Probe acuity epoch extraction. Epochs are extracted relative to the probe stimulation for experimental trials and relative to the cue for the control trials. Epoch extraction and prepossessing is conducted according to \ref{sec:DecReadInfoCodeMeth} before trials are passed to the logistic regression model. Predictions $y_{pred}$ are output as labels identifying the intput as either originating from a control cue stimulation or experimental probe stimulation with a given delay relative to the cue stimulation.}
    \label{fig:ProbeProbeSelection}
\end{figure}

To explore probe to probe confusion and thus the acuity of the separation ability of the networks we utilized the same epoch extraction method described in section \ref{sec:UpMem} for both experiment 1 and 2. Epoch selection example for experiment 2 is given in Figure: \ref{fig:ProbeProbeSelection}
We then constructed two multiclass classification tasks, one for experiment 1 and one for experiment 2 where the classes corresponded to each of the experimental conditions and the control condition. The resulting confusion matrices allow us to identify whether any probe responses are confused at what time, which would
indicate the acuity of the separation property. 

\subsubsection{Results}
\label{sec:PPAcuRes}

We observed that performance was generally lower in the multiclass task than in the
binary task.
The results from experiment 1 were generally good, showing little confusion between conditions with the exception of between the control cue and the 7000 ms probe delay. However, network A3 showed low accuracy, while B3 was close to chance level. Results for the best parameter combination is given in Table \ref{ap:ppAexp1} in the appendix.

The confusion matrices from experiment 2 instead showed an increasing confusion between close probes as the probe delay increased, particularly between the 600
and 800 ms probes, see Figures \ref{fig:NetAMulticalssExp2}, \ref{fig:NetBMulticalssExp2}, and Table and \ref{ap:ppAexp2} in the appendix. The confusion with the cue stimulation did however stay low.

Note that the accuracy of the 800 ms probe and the control cue is not entirely comparable to the rest of the conditions because they only have one
”close” condition, i.e probe delays that are within 200 ms of them, while the other ones have two.

\subsubsection{Discussion}
\label{sec:PPAcuDis}
Our results indicate that the networks encode information with higher acuity when stimulations are input with close temporal distance and show a gradual decay as the distance between cue and probe is increased. Given the low temporal consistency of spikes in later phases of evoked spike responses this may be unsurprising. However, we do see high performance at 100 and 200 ms in most cases and 400 ms in some cases, indicating that we may still gain high acuity information from the phases with inconsistent spike timing.

\newpage 

\begin{figure}[!htb]
    \centering{}
    \includegraphics[width = 0.8\linewidth]{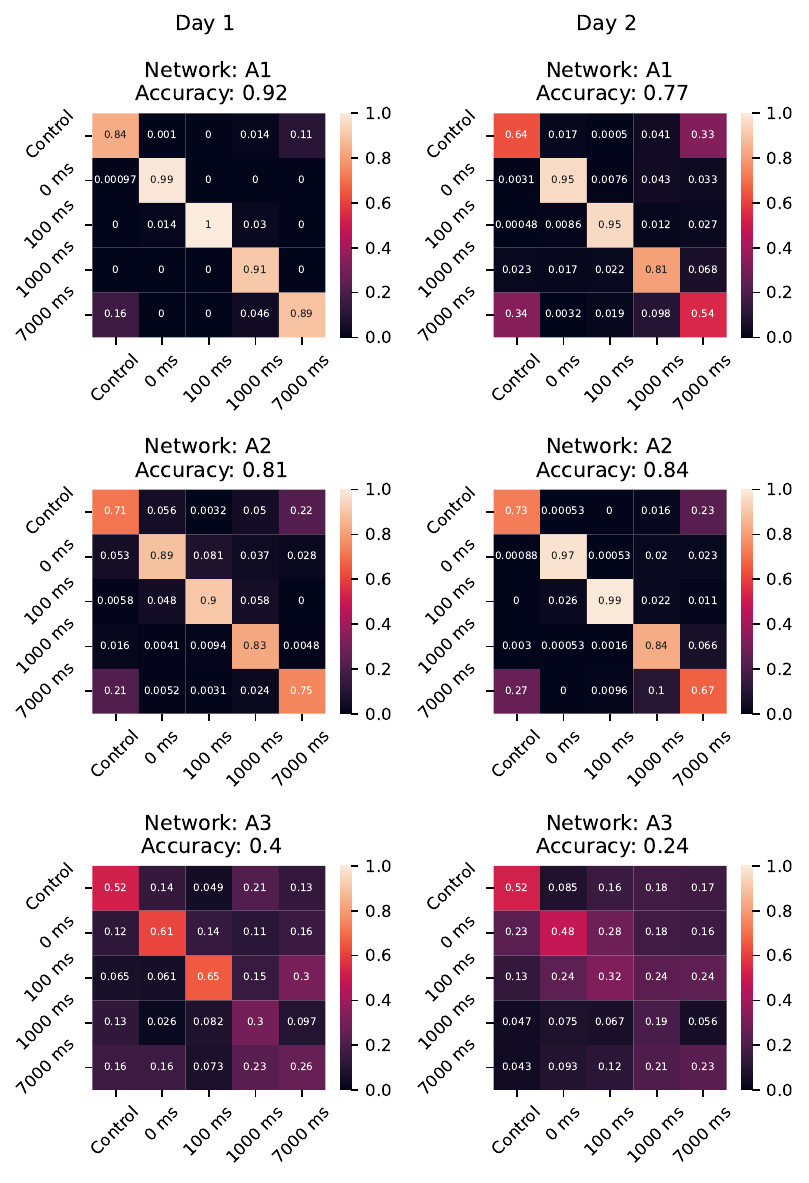}
    \caption{Confusion matrices for networks A1,A2 and A3 for classification between all conditions in experiment 1. We can here see that the performance is generally good, with some confusion appearing between the control and 7000 ms probe for most networks while A3 shows overall poor performance.}
    \label{fig:NetAMulticalssExp1}
\end{figure}

\newpage 
\begin{figure}[!htb]
    \centering{}
    \includegraphics[width = 0.8\linewidth]{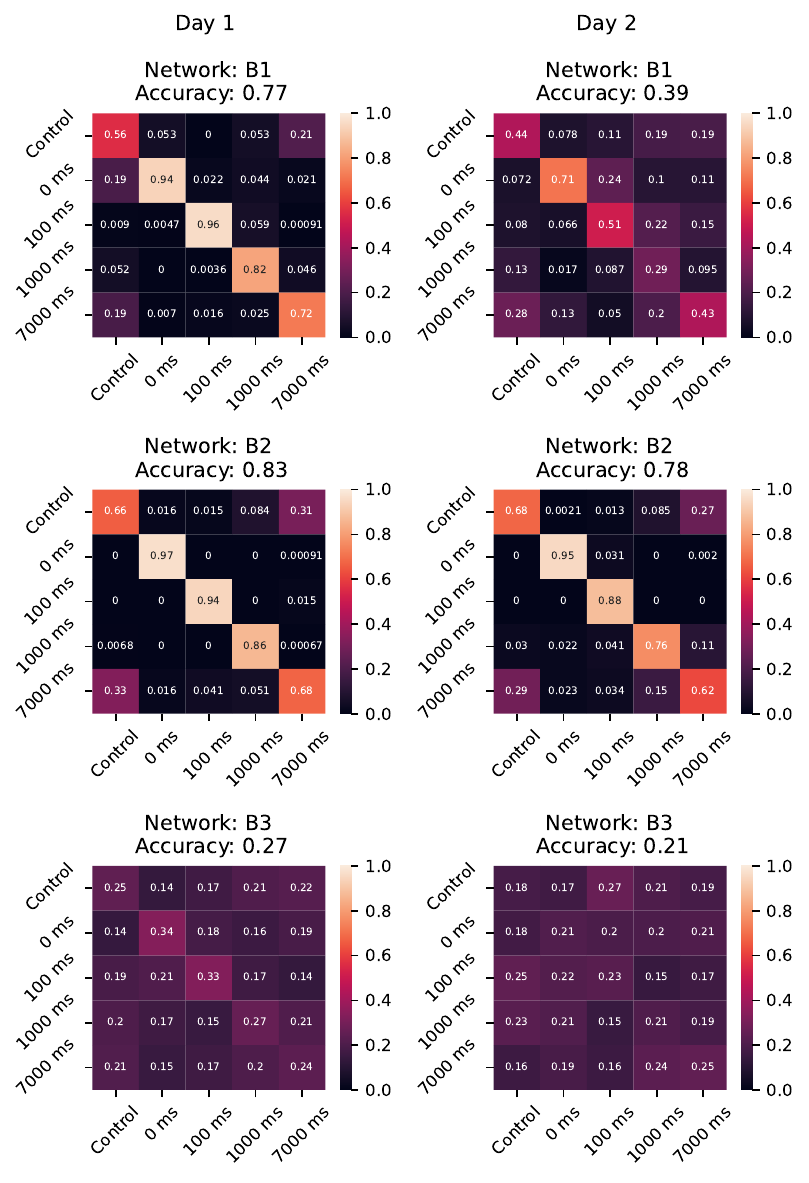}
    \caption{Confusion matrices for networks B1,B2 and B3 for classification between all conditions in experiment 1. The confusion matrices show a similar pattern to the A networks, but we here also see a more drastic decrease in performance over time for B1.}
    \label{fig:NetBMulticalssExp1}
\end{figure}

\newpage 

\begin{figure}[!htb]
    \centering{}
    \includegraphics[width = 0.8\linewidth]{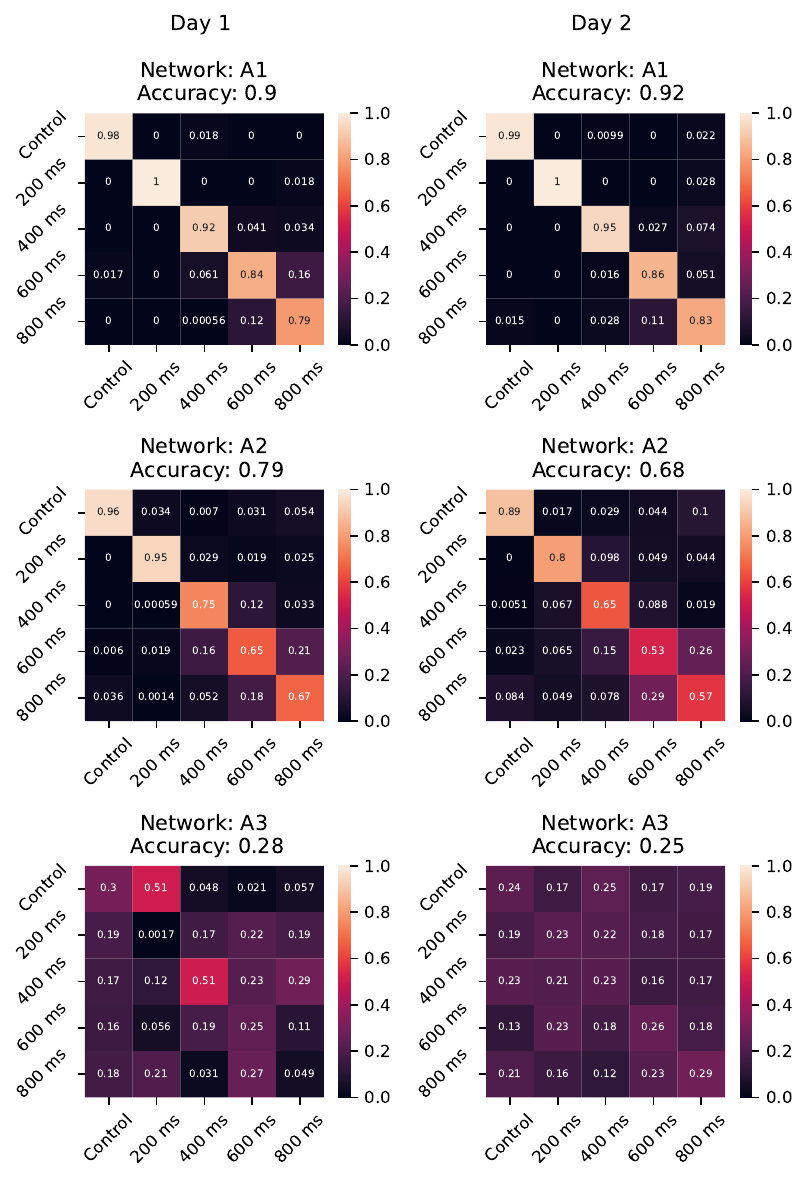}
    \caption{Confusion matrices for networks A1,A2 and A3 for classification between all conditions in experiment 2. Network A1 shows overall excellent performance, while A2 shows increased confusion between later probes }
    \label{fig:NetAMulticalssExp2}
\end{figure}

\newpage
\begin{figure}[!htb]
    \centering{}
    \includegraphics[width = 0.8\linewidth]{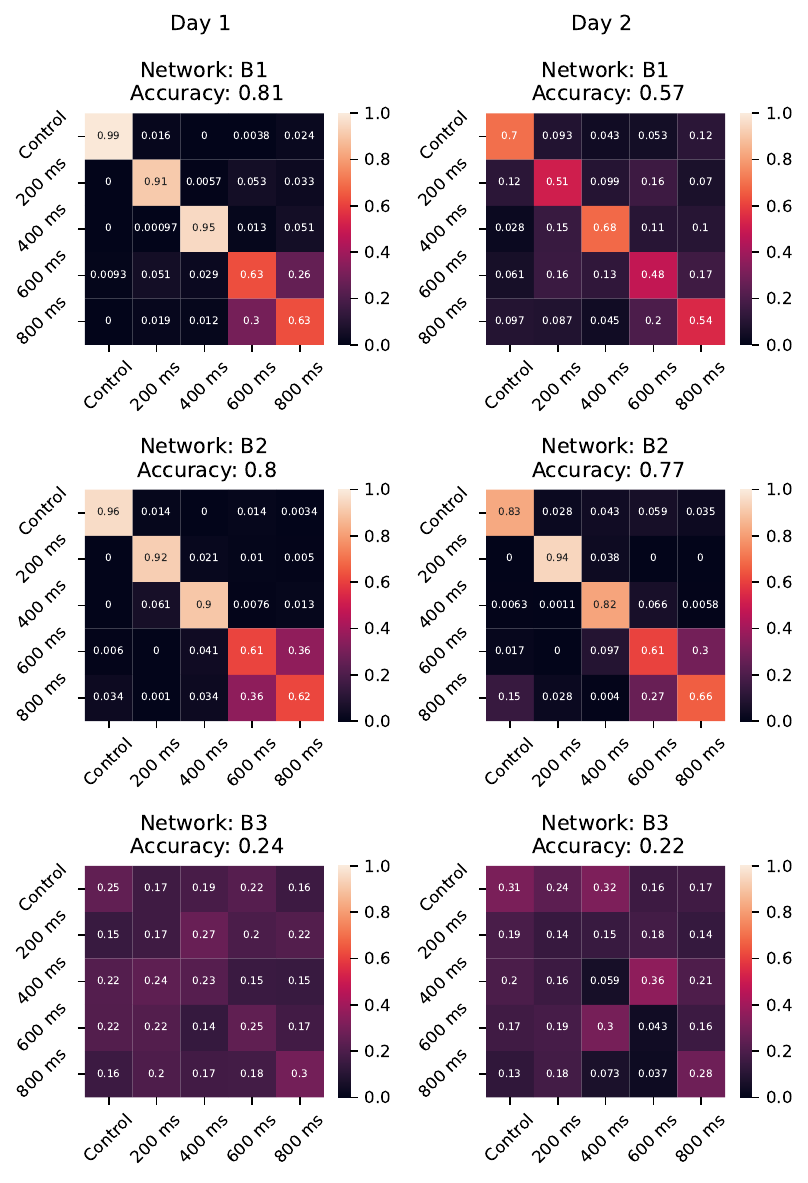}
    \caption{Confusion matrices for networks B1,B2 and B3 for classification between all conditions in experiment 2. The confusion patterns here shows a similar increase in confusion for later probes for B1 on day 1 and B2 for both days. B1 on day 2 again shows a more drastic decrease in performance while B3 performs overall poorly.}
    \label{fig:NetBMulticalssExp2}
\end{figure}

\newpage

\section{Summary Discussion}
\label{sec:SumDis}

Our main findings show that in-vitro neural networks can have an upper separability bound as large as 7 seconds, but with decreased accuracy at larger delays, and a lower separability bound as low as 0 (36) ms, but with relatively low accuracy and often very short term effects on the overall trajectory of the networks. Their acuity can furthermore be sufficient to distinguish between timing differences of 100 (136) ms, but this acuity decreases with increasing delays between stimulation timing. The networks can encode information concerning the timing between two stimulation pulses both in specific spike times and in the rate of the spike response and these show different performance over different probe delays and phases of the response. 
The performance of the networks was however not stable over time, with large differences in separability being observed between day one and two of our experiments in certain cases. 

These results suggests that in-vitro neural networks can encode information input using the stimulation timing method. While stimulation timing up to 7 seconds can be encoded into the activity of the networks, this appears to come at the price of accuracy of separability. Due to this issue and the acuity issue, encoding information within a 0-400 (36-436)ms time frame appears to provide best performance and may thus be used as a guideline for further exploration into in-vitro neural networks as reservoir computing systems. 

The short effect of the probe stimulation on the evoked response does however pose a challenge. To be able to compute an output based on multiple inputs the effect of all inputs must temporally overlap otherwise they cannot affect each other. The short effect of the stimulations on the evoked response suggests that the time window within which multiple stimulations can elicit overlapping responses is small, which may necessitate using input with shorter time delays than the 0-400 (36-436) ms suggestion. This could potentially be solved by tuning the networks to other phases were their response does not encompass the entire network in the form of network wide barrages as done in \cite{heiney2019assessment, wagenaar2005controlling}. 
However, if the networks use neural exhaustion to form long, fading, memories, the upper bound of separability may be drastically decreased by such tuning as the network would no longer have large exhausting bursts available to store such memories. However, given the decreased acuity for timings of 600-800 ms, timings exceeding these values may not be useful for high accuracy computation and thus may be sacrificed without major decreases in computational power. 

Concerning the Decoder-readout parameters, the different performance, depending on time-bin size, we saw under certain network-condition combinations indicates that using one singular time-bin size may not be optimal. Instead we may extract multiple time-bin sizes over the same epoch size to capture information stored at multiple scales. However, while doing so we may need to consider overfitting issues given the already high dimensionality of the epochs when using small time bin sizes. Furthermore, it may be of interest to assess if different time-bin sizes are optimal for different tasks on the same input, given that we may attempt to solve multiples problems on the same datastream in parallel.

The instability of the performance between days associated with the metabolic demands of the networks relative to media replenishment does also present a challenge in the application of in-vitro neural reservoirs. To maintain a continuous supply of nutrients, a possible solution may lie in using systems that allow for continuous nutrient exchange using microfluidics (for a review of applications of microfluidics see: \cite{wang2009microfluidics}, for specific example for in-vitro neural networks on MEA see: \cite{heo2016dependence}) to ensure stable conditions. However, other features of biological networks may be at play. Representational drift can cause the neural representations to change over time \cite{rule2019causes} which may cause confusion in the decoder as the representational drift causes the need for re-training. 

The different behavior between the different networks and the presumably large difference in network architecture that arises due to the self organization process implicit in-vitro neural networks also align with the pros and cons of \textit{(im)mortal computation} as described by Hinton \cite{hinton2022forward}. Typical AI systems in deployment utilizes their \textit{immortality}, in the sense that the weights of the model are separate from the hardware and can thus be copied to an arbitrary number of instances at will to dynamically scale up and down according to demand. This benefit of \textit{immortality} works well with cloud infrastructures and is the backbone of how many AI applications operate. Biological networks on the other hand, benefit from the energy efficiency of \textit{mortal} biological neurons but one cannot perform similar dynamical scaling because software and hardware are not separated. Furthermore, due to the uniqueness of each individual network, the weights learned for one network and task cannot be generalized to another network solving the same task. This potential trade-off may limit some of the application areas of biological computation. This means that the application of biological neural networks for computing tasks may be easier in other areas, such as low energy sensing devices \cite{hernandez2022high}, control systems \cite{kagan2022vitro, demarse2005adaptive, tessadori2012modular, karniel2005computational, aaser2017towards, masumori2018autonomous} and similar types of computing that does not require dynamic scaling.

\bigskip
In conclusion we have provided an in-depth exploration of key features of neural computation in in-vitro neural networks on inputs encoded through stimulation timing. Namely, we have established a lower bound, and upper bound and identified a current bound on acuity of linear separability of this encoding method as well as described some limitations of these features. Our findings may provide guidelines for other researchers exploring these topics and may accelerate the search for optimal encoder settings by reducing the search space. As we also show that the encoded information may be represented in different forms, e.g. spike time and rate, across the evoked response we also contribute insight into the construction of decoders as these must account for the different coding formats to optimally extract the result of computations from the system.

\subsection{Limitations}
\label{sec:Lim}
Our networks showed large bursts when stimulated which may not be optimal for information processing. This relates to the phase of the networks as they can develop into critical or non-critical phases \cite{pasquale2008self}. Our results may thus not generalize to networks in other phases than the ones our network were in. This may be particularly important for more extensive stimulation sequences as the large bursts we saw in our networks may make it difficult to input longer sequences of stimulations, something that would be essential if any useful computations is to be done with these systems.
We also did not fully establish the bound on acuity in our study, as we have only tested variations down to 100 (136) ms. Importantly this could be affected by the phase of the network as tuning networks to criticality has been shown to improve dynamic range for variations in current amplitude \cite{shew2009neuronal} which is equivalent to improving acuity. 

We furthermore do not compare our encoding method to other methods. While we provide theoretical arguments for why stimulation timing may be the optimal encoding method, it may be possible that there are unforeseen benefits to alternative methods, or challenges to the current method that voids the suggested benefits.

Finally, given that we used counterbalancing over the two days each experiment was ran, we cannot entirely disentangle the effect of time since feeding from a potential order effect. However, given the decreased accuracy at 7 seconds, it would appear unlikely that fading memories would be stored for the 30 seconds rest period. Therefor the order of the conditions is not likely to have affected the outcomes.

\subsection{Future Work}
\label{sec:FutWork}
Future studies should systematically explore different encoding methods, by testing for relevant features like the bounds and acuity of the method to be able to analyse the trade-offs between different approaches. Furthermore, the phase of the networks should be varied to explore how the dynamics affects their information processing under different encoding schemes. An additional step to this line of research should involve exploring the interaction between multiple encoded pieces of information as well as explore how learning in the networks affect their processing capabilities. In addition, the stability of the network responses should be tested over long time-period to assess if representational drift may become an issue for the decoder and how to deal with these appropriately. Ideally all such systems should be capable of ensuring stable conditions for the neurons to avoid the confounding variable of nutrient instability and similar issues.

\section{Acknowledgements}
This work was partially financed by the Research Council of Norway's DeepCA project, grant agreement 286558.

\section{Contributor Role Taxonomy}
\credit{Trym A. E. Lindell}{0.4 ,0.7 ,1   ,0     ,0  ,0.6   ,0.25   ,0      ,1  ,0      ,0.6    ,0.9    ,0.9  ,0.5}
\credit{Ola H. Ramstad}    {0.2 ,0.3 ,0   ,0     ,1  ,0.2   ,0.25   ,0      ,0  ,0      ,0.1    ,0.1    ,0.1  ,0.2}
\credit{Ioanna Sandvig}     {0   ,0   ,0   ,0.2   ,0  ,0     ,0.25   ,0.3    ,0  ,0.2    ,0      ,0      ,0    ,0.05}
\credit{Axel Sandvig}      {0   ,0   ,0   ,0.2   ,0  ,0     ,0      ,0.3    ,0  ,0.1    ,0      ,0      ,0    ,0.05}
\credit{Stefano Nichele}   {0.4 ,0   ,0   ,0.6   ,0  ,0.2   ,0.25   ,0.3    ,0  ,0.7    ,0.2    ,0      ,0    ,0.2}

\insertcredits

\bibliography{ijuc}

\section{Appendix}





\newpage
\appendix 

\newpage
\begin{table}[htb!]
\begin{center}
\caption{Upper Separability Bound - Day 1 - Highest Performing Parameters for Epoch Size x Time Bin Size x Offset}
\label{ap:upSepD1}
\begin{tabular}{lllrrr}
\toprule
Network & Condition & Epoch Length &  Time-Bin Size &  Offset &  Accuracy \\
\midrule
  A1 &      0 ms &           64 &             32 &     115 &  1.000000 \\
  A1 &    100 ms &         1024 &              8 &       1 &  1.000000 \\
  A1 &   1000 ms &            2 &              2 &      89 &  0.997500 \\
  A1 &   7000 ms &          512 &             64 &      74 &  0.851000 \\
  A2 &      0 ms &         1024 &             64 &      27 &  0.904167 \\
  A2 &    100 ms &          128 &            128 &     156 &  0.969167 \\
  A2 &   1000 ms &            2 &              2 &      92 &  0.980667 \\
  A2 &   7000 ms &           16 &              8 &       3 &  0.716333 \\
  A3 &      0 ms &         1024 &            128 &       1 &  0.705833 \\
  A3 &    100 ms &           32 &              8 &     106 &  0.684000 \\
  A3 &   1000 ms &          128 &            128 &      12 &  0.662000 \\
  A3 &   7000 ms &            2 &              1 &     395 &  0.597333 \\
  B1 &      0 ms &            4 &              4 &      53 &  0.843000 \\
  B1 &    100 ms &         1024 &            512 &      47 &  0.961667 \\
  B1 &   1000 ms &            8 &              1 &     151 &  0.912167 \\
  B1 &   7000 ms &          512 &            128 &     163 &  0.784333 \\
  B2 &      0 ms &           64 &             64 &      46 &  0.991167 \\
  B2 &    100 ms &            2 &              2 &      10 &  1.000000 \\
  B2 &   1000 ms &          512 &             16 &       5 &  0.959000 \\
  B2 &   7000 ms &          256 &             64 &     195 &  0.687833 \\
  B3 &      0 ms &            1 &              1 &       0 &  0.500000 \\
  B3 &    100 ms &            1 &              1 &       0 &  0.500000 \\
  B3 &   1000 ms &            1 &              1 &       0 &  0.500000 \\
  B3 &   7000 ms &            1 &              1 &       0 &  0.500000 \\
\bottomrule
\end{tabular}
\end{center}

\end{table}

\newpage 
\begin{table}[htb!]

\begin{center}
\caption{Upper Separability Bound - Day 2 - Highest Performing Parameters for Epoch Size x Time Bin Size x Offset}
\label{ap:upSepD2}
\begin{tabular}{lllrrr}
\toprule
Network & Condition & Epoch Length &  Time-Bin Size &  Offset &  Accuracy \\
\midrule
  A1 &      0 ms &          256 &             64 &       1 &  0.963167 \\
  A1 &    100 ms &          128 &             64 &      16 &  0.990667 \\
  A1 &   1000 ms &           64 &             64 &     510 &  0.975833 \\
  A1 &   7000 ms &          512 &            128 &     661 &  0.681500 \\
  A2 &      0 ms &           64 &             32 &       9 &  0.979500 \\
  A2 &    100 ms &          512 &            128 &       0 &  0.992167 \\
  A2 &   1000 ms &           32 &              2 &       7 &  0.982667 \\
  A2 &   7000 ms &          128 &            128 &     543 &  0.757333 \\
  A3 &      0 ms &            2 &              1 &     442 &  0.538333 \\
  A3 &    100 ms &            8 &              8 &     482 &  0.552500 \\
  A3 &   1000 ms &           64 &             64 &     344 &  0.564500 \\
  A3 &   7000 ms &          128 &              4 &     288 &  0.559667 \\
  B1 &      0 ms &           64 &              8 &       0 &  0.654500 \\
  B1 &    100 ms &            4 &              4 &    1516 &  0.586833 \\
  B1 &   1000 ms &           16 &             16 &     703 &  0.606833 \\
  B1 &   7000 ms &           16 &             16 &      51 &  0.552833 \\
  B2 &      0 ms &          512 &            128 &      15 &  0.988667 \\
  B2 &    100 ms &            2 &              2 &       0 &  1.000000 \\
  B2 &   1000 ms &          512 &             32 &      10 &  0.921667 \\
  B2 &   7000 ms &           32 &             16 &     157 &  0.689500 \\
  B3 &      0 ms &            1 &              1 &       0 &  0.500000 \\
  B3 &    100 ms &            1 &              1 &       0 &  0.500000 \\
  B3 &   1000 ms &            1 &              1 &       0 &  0.500000 \\
  B3 &   7000 ms &            1 &              1 &       0 &  0.500000 \\
\bottomrule
\end{tabular}
\end{center}

\end{table}

\begin{table}[htb!]
\begin{center}
\caption{Lower Separability Bound - Day 1 -  Highest Performing Parameters for Epoch Size x Time Bin Size x Offset}
\label{ap:lowSepD1}
\begin{tabular}{lllrrr}
\toprule
Network &   Condition & Epoch Length &  Time-Bin Size &  Offsett &  Accuracy \\
\midrule
A1 & cue 0 ms & 64 & 16 & 1015 & 0.691944 \\
A1 & cue 100 ms & 1024 & 4 & 0 & 0.750278 \\
A1 & cue 1000 ms & 64 & 64 & 1 & 0.900833 \\
A1 & cue 7000 ms & 256 & 256 & 2 & 0.995000 \\
A2 & cue 0 ms & 128 & 16 & 115 & 0.725833 \\
A2 & cue 100 ms & 64 & 4 & 42 & 0.643333 \\
A2 & cue 1000 ms & 1024 & 256 & 746 & 0.613611 \\
A2 & cue 7000 ms & 2 & 1 & 121 & 0.992778 \\
A3 & cue 0 ms & 8 & 8 & 304 & 0.676389 \\
A3 & cue 100 ms & 32 & 16 & 29 & 0.621111 \\
A3 & cue 1000 ms & 1024 & 32 & 1279 & 0.542500 \\
A3 & cue 7000 ms & 64 & 64 & 2 & 0.601389 \\
B1 & cue 0 ms & 16 & 16 & 0 & 0.695000 \\
B1 & cue 100 ms & 32 & 2 & 391 & 0.646111 \\
B1 & cue 1000 ms & 128 & 8 & 19 & 0.675278 \\
B1 & cue 7000 ms & 256 & 256 & 190 & 0.791111 \\
B2 & cue 0 ms & 256 & 16 & 175 & 0.655278 \\
B2 & cue 100 ms & 128 & 8 & 0 & 0.788611 \\
B2 & cue 1000 ms & 1 & 1 & 1 & 1.000000 \\
B2 & cue 7000 ms & 32 & 32 & 81 & 0.977500 \\
B3 & cue 0 ms & 1 & 1 & 0 & 0.500000 \\
B3 & cue 100 ms & 1 & 1 & 0 & 0.500000 \\
B3 & cue 1000 ms & 1 & 1 & 0 & 0.500000 \\
B3 & cue 7000 ms & 1 & 1 & 0 & 0.500000 \\
\bottomrule
\end{tabular}
\end{center}

\end{table}

\begin{table}[htb!]
\begin{center}
\caption{Lower Separability Bound - Day 2 -  Highest Performing Parameters for Epoch Size x Time Bin Size x Offset}
\label{ap:lowSepD2}
\begin{tabular}{lllrrr}
\toprule
Network &   Condition & Epoch Length &  Time-Bin Size &  Offset &  Accuracy \\
\midrule
A1 & cue 0 ms & 64 & 16 & 1791 & 0.695556 \\
A1 & cue 100 ms & 32 & 16 & 0 & 0.745000 \\
A1 & cue 1000 ms & 8 & 4 & 56 & 0.761111 \\
A1 & cue 7000 ms & 4 & 1 & 54 & 0.970000 \\
A2 & cue 0 ms & 256 & 256 & 2031 & 0.676667 \\
A2 & cue 100 ms & 128 & 16 & 4 & 0.706944 \\
A2 & cue 1000 ms & 32 & 32 & 6 & 0.945833 \\
A2 & cue 7000 ms & 64 & 32 & 18 & 0.988611 \\
A3 & cue 0 ms & 2 & 2 & 419 & 0.546111 \\
A3 & cue 100 ms & 4 & 4 & 334 & 0.550556 \\
A3 & cue 1000 ms & 16 & 16 & 1815 & 0.538333 \\
A3 & cue 7000 ms & 128 & 16 & 1496 & 0.554444 \\
B1 & cue 0 ms & 16 & 16 & 4 & 0.704722 \\
B1 & cue 100 ms & 16 & 8 & 371 & 0.594722 \\
B1 & cue 1000 ms & 256 & 64 & 751 & 0.597222 \\
B1 & cue 7000 ms & 512 & 8 & 0 & 0.678056 \\
B2 & cue 0 ms & 4 & 4 & 117 & 0.632222 \\
B2 & cue 100 ms & 32 & 16 & 0 & 0.791389 \\
B2 & cue 1000 ms & 2 & 1 & 49 & 0.965278 \\
B2 & cue 7000 ms & 16 & 16 & 6 & 0.991667 \\
B3 & cue 0 ms & 1 & 1 & 0 & 0.500000 \\
B3 & cue 100 ms & 1 & 1 & 0 & 0.500000 \\
B3 & cue 1000 ms & 1 & 1 & 0 & 0.500000 \\
B3 & cue 7000 ms & 1 & 1 & 0 & 0.500000 \\
\bottomrule
\end{tabular}
\end{center}

\end{table}

\begin{table}[htb!]
\begin{center}
\caption{Probe - Probe Acuity (experiment 1) - Highest Performing Parameters for Epoch Size x Time Bin Size x Offset}
\label{ap:ppAexp1}
\begin{tabular}{lllrrr}
\toprule
Network &   Day & Epoch Length &  Time Bin Size &  Offset &  Accuracy \\
\midrule
  A1 & Day 1 &          512 &             16 &      46 &  0.922000 \\
  A1 & Day 2 &          512 &             16 &       7 &  0.772000 \\
  A2 & Day 1 &           64 &              4 &      15 &  0.808556 \\
  A2 & Day 2 &          256 &             16 &       1 &  0.835778 \\
  A3 & Day 1 &          512 &            128 &       8 &  0.402667 \\
  A3 & Day 2 &         1024 &              4 &      24 &  0.240111 \\
  B1 & Day 1 &          512 &            128 &       5 &  0.770889 \\
  B1 & Day 2 &           64 &              1 &      21 &  0.390556 \\
  B2 & Day 1 &         1024 &             16 &      38 &  0.829889 \\
  B2 & Day 2 &          512 &             32 &       1 &  0.783556 \\
  B3 & Day 1 &           32 &              4 &       4 &  0.270778 \\
  B3 & Day 2 &            1 &              1 &       2 &  0.213222 \\
\bottomrule
\end{tabular}
\end{center}

\end{table}

\begin{table}[htb!]
\begin{center}
\caption{Probe - Probe Acuity (experiment 2) - Highest Performing Parameters for Epoch Size x Time Bin Size x Offset}
\label{ap:ppAexp2}
\begin{tabular}{lllrrr}
\toprule
Network &   Day & Epoch Length &  Time Bin Size &  Offset &  Accuracy \\
\midrule
  A1 & Day 1 &          256 &             32 &       8 &  0.902222 \\
  A1 & Day 2 &          256 &             32 &       3 &  0.917333 \\
  A2 & Day 1 &          512 &             64 &       5 &  0.788222 \\
  A2 & Day 2 &           64 &              1 &      19 &  0.681556 \\
  A3 & Day 1 &            1 &              1 &       4 &  0.280556 \\
  A3 & Day 2 &            1 &              1 &       4 &  0.248111 \\
  B1 & Day 1 &          512 &              2 &      12 &  0.813667 \\
  B1 & Day 2 &          512 &              4 &       7 &  0.566556 \\
  B2 & Day 1 &         1024 &            128 &       0 &  0.804778 \\
  B2 & Day 2 &          256 &             32 &       5 &  0.772111 \\
  B3 & Day 1 &            8 &              4 &       0 &  0.235889 \\
  B3 & Day 2 &            2 &              1 &       4 &  0.216667 \\
\bottomrule
\end{tabular}
\end{center}

\end{table}

\end{document}